\documentclass[journal]{IEEEtran}

\usepackage{amssymb, amsmath, graphicx, cite, color, bm, bbm}
\usepackage[caption=false,font=footnotesize]{subfig}

\newtheorem{thm}{Theorem}
\newtheorem{cor}{Corollary}
\newtheorem{lem}{Lemma}
\newtheorem{prop}{Proposition}

\newcommand{\BV}{{\sf BV}}
\newcommand{\TV}{{\sf TV}}
\newcommand{\wv}{{\sf wv}}
\newcommand{\Proj}{{\sf Proj}}
\newcommand{\sign}{{\sf sign}}
\newcommand{\wh}[1]{\widehat{#1}}
\newcommand{\info}{{\sf I}}
\newcommand{\ener}{{\sf E}}
\newcommand{\maxx}{{\sf max}}

\title{Information and Energy Transmission with Wavelet-Reconstructed Harvesting Functions }

\author{Daewon Seo and Yongjune Kim
\thanks{ This paper was presented in part at the 2022 IEEE International Symposium on Information Theory (ISIT) \cite{SeoK2022_isit}. (Corresponding author: Yongjune Kim)}
\thanks{ This work was supported in part by DGIST Start-up Fund Program of the Ministry of Science and ICT (No.~2023010048) and the National Research Foundation of Korea (NRF) grant funded by the Korea government (MSIT) (No.~RS-2023-00212103).}
\thanks{ D.~Seo is with the Department of Electrical Engineering and Computer Science, DGIST, Daegu 42988, South Korea (e-mail: dwseo@dgist.ac.kr). Y.~Kim is with the Department of Electrical Engineering, Pohang University of Science and Technology (POSTECH), Pohang 37673, South Korea, and also with the Institute for Convergence Research and Education in Advanced Technology, Yonsei University, Seoul 03722, South Korea (e-mail: yongjune@postech.ac.kr). }
}

\begin{document}
\maketitle

\begin{abstract}
In practical simultaneous information and energy transmission (SIET), the exact energy harvesting function is usually unavailable because an energy harvesting circuit is nonlinear and nonideal. In this work, we consider a SIET problem where the harvesting function is accessible only at experimentally-taken sample points and study how close we can design SIET to the optimal system with such sampled knowledge. Assuming that the harvesting function is of bounded variation that may have discontinuities, we separately consider two settings where samples are taken without and with additive noise. For these settings, we propose to design a SIET system as if a wavelet-reconstructed harvesting function is the true one and study its asymptotic performance loss of energy and information delivery from the true optimal one. Specifically, for noiseless samples, it is shown that designing SIET as if the wavelet-reconstructed harvesting function is the truth incurs asymptotically vanishing energy and information delivery loss with the number of samples. For noisy samples, we propose to reconstruct wavelet coefficients via soft-thresholding estimation. Then, we not only obtain similar asymptotic losses to the noiseless case but also show that the energy loss by wavelets is asymptotically optimal up to a logarithmic factor.
\end{abstract}

\begin{IEEEkeywords}
energy harvesting, capacity-energy function, wavelets, functions of bounded variation
\end{IEEEkeywords}

\section{Introduction} \label{sec:intro}
Simultaneous information and energy transmission (SIET) has been adopted for a wide variety of applications such as radio-frequency identification (RFID) tags, wireless remote sensors, as well as in-vivo biomedical devices. Such wide applications have initiated extensive research on SIET systems. Since the first study of fundamental limits of SIET \cite{Varshney2008}, there have been a lot of subsequent works, for instance, SIET for frequency-selective additive white Gaussian noise (AWGN) channels \cite{GroverS2010}, finite block-length setting \cite{ZuhraPPA2022}, wireline settings \cite{Varshney2012}, and wireless settings \cite{ZhangMH2015, ClerckxZSNKP2019, ZhouZH2013, Gastpar2007, BanawanU2016, AmorPKP2017, AlevizosB2018, WuC2022}. Also, see \cite{ClerckxKCK2022} for a recent survey. The wireless problems are particularly referred to as SWIPT (simultaneous wireless information and power transmission) for short.

One of the major challenges for SIET and SWIPT is that energy harvesting functions in practice do not follow simple energy laws from physics, such as $\int y^2(t) dt$ for received signal $y(t)$. Moreover, harvesting circuits' nonlinearity and nonideality, such as nonlinear components, imperfect impedance matching, nonideal signal filters, and parasitic capacitors, generally make the harvesting function complicated and intractable. It is then impossible to design exactly optimal SIET systems as 
the exact knowledge about the true harvesting function is not available. 

To deal with SIET having such intractable harvesting functions, there are two dominant approaches in literature. The first approach splits a harvesting circuit into multiple components, approximates each, e.g., using Taylor approximation, and then combines them to obtain an end-to-end harvesting function in closed form \cite{ValentaD2014, BoshkovskaNZS2015, SoyataCHsoyata2016, ClerckxB2016, VarastehRC2017, KangKK2018}. In particular in \cite{ClerckxB2016}, a nonlinear analytic model for a harvesting circuit is derived by approximating an ideal diode's characteristics via the Taylor series, and based on the model, it proposes to use multisine waveforms to deliver the largest energy. However, this approach has limitations because individual components' laws may not even follow ideal circuit laws due to coupling effect, parasitic capacitors, etc., and then such approximation errors could accumulate. For example, the parasitic components resulting from interrelationships make it difficult to analytically describe an equivalent end-to-end model for a circuit containing diodes.

Indeed, the true end-to-end energy harvesting function may only be available through experimental samples or circuit simulations \cite{LeMF2008, NintanavongsaMLR2012, StoopmanKVPS2013, StoopmanKVPS2014, SamplePSS2013, Zhang_et_al2019}. Inspired by this, recent literature attempts to take experimental samples and use a deep learning approach. Reference \cite{VarastehHC2020} first reconstructs a harvesting function from samples by deep learning regression, and then, it optimizes constellation points for modulation by autoencoders based on neural networks. Reference \cite{ShaninCS2021} focuses on the memory effect of harvesting circuits. To capture the memory effect, it proposes a Markov decision process (MDP) model for the harvesting circuit and uses neural networks to estimate the MDP's parameters from data created by a circuit simulator. Despite their success in practice, the deep learning approach has a critical drawback in that it generally provides no theoretical performance guarantees. In addition, since optimization via deep learning is a black-box algorithm, it generally provides limited insights into SIET. For instance, it cannot answer how many experimental samples are needed to attain a particular performance or how close our SIET design is to the optimal SIET.

Our previous work \cite{SeoV2019} accomplishes the learning approach and theoretical guarantees simultaneously by using splines and local polynomials. The work is motivated by the fact that for low-dimensional functions, classic function interpolation methods perform as good as deep learning regression but also provide theoretical guarantees. Specifically, assuming that a smooth harvesting function---formally, belonging to a class of Sobolev functions---is known only at sample points, it characterizes asymptotic performance loss of information rate and energy incurred by such sampled knowledge \cite{SeoV2019}. Instead of deep learning regression, it suggests using the spline and local polynomial methods to reconstruct the unknown harvesting function and using it as a proxy for the true one. The results are promising as performance losses from the true optimal performance are asymptotically negligible with the number of samples, but the smoothness assumption is restrictive because the harvesting function occasionally varies very quickly as a signal varies, e.g., due to diode's mode transition, hardware defects, or unexpected resonance of circuits if frequency modulation is adopted. Then, collected real data displays a sharp change in the true harvesting function, e.g., Figure 4 in \cite{VarastehHC2020}. Even when the true function is (formally Sobolev) smooth, such sharp change implies a large magnitude of its derivative, which is in turn captured by a hidden constant factor of the loss characterization that depends on derivatives \cite{SeoV2019}. Therefore, the resulting loss bounds are vacuous. In such a case, it is more reasonable to model it as a function of bounded variation that may have discontinuities, but its total variation is bounded. This is a larger class of functions than the Sobolev functions, for which the spline reconstruction method does not perform well.

This work considers a class of harvesting functions of bounded variation, which may have a finite number of jumps (discontinuities). In addition, we assume that the harvesting function is accessible only at discrete sample points, either in the absence of noise or in the presence of noise. In these settings, our ultimate goal is to discuss how to design information-theoretic \mbox{(near-)} optimal SIET systems and how they perform. To this end, our proposed method first reconstructs the harvesting function using wavelet methods and then designs a transmission codebook based on it. Unlike the deep learning-based approach~\cite{VarastehHC2020}, using well-studied wavelet methods enables us to characterize its performance loss analytically. Also, this work generalizes our previous work \cite{SeoV2019} that uses splines and local polynomials to reconstruct only smooth functions. Wavelets are clearly different from splines and local polynomials, and using wavelets is generally minimax (near-) optimal for a broader class of functions. The results show that the losses of energy and information rate asymptotically vanish with the number of samples. To be specific, our findings are summarized as follows.
\begin{itemize}
	\item For noiseless samples, if SIET is designed based on a wavelet-reconstructed harvesting function, the energy delivery can be asymptotically close to that with a fully known harvesting function. The information transmission can also be asymptotically close to the optimal one in the interior of the energy domain.
	
	\item For noisy samples, we propose to reconstruct the harvesting function by wavelets with soft-thresholding coefficient estimation. Then, the wavelet-based SIET results in the expected losses of energy and information that asymptotically vanish with the number of samples. Moreover, we show that the energy loss due to the sampled knowledge is asymptotically optimal up to a logarithmic factor for a certain channel, i.e., the loss cannot be further reduced.
\end{itemize}

The remainder of this paper is organized as follows. Section \ref{sec:formulation} formalizes the problem and introduces the basics of wavelets. Section \ref{sec:noiseless_samples} investigates the energy and information losses when samples are taken without noise. Section \ref{sec:noisy_samples} investigates the expected energy and information losses when samples are taken with noise. Section \ref{sec:evaluation} numerically illustrates the proposed scheme and compares it with spline-based \cite{SeoV2019} and deep learning-based ones \cite{VarastehHC2020}. Finally, Section \ref{sec:conclusion} concludes the paper.

\section{Background and Formulation} \label{sec:formulation}
\subsection{Capacity-Energy Function}
The goal of SIET is to deliver information and energy simultaneously over a noisy channel. In this work, we particularly assume that the harvesting function is known only at discrete sample points, for example, through measurements at those points. The codeword can be also thought of as a sequence of signals or waveforms. We take an information-theoretic approach \cite{CoverT1991} to the problem as it provides a rigorous theoretical framework for analyzing the fundamental limits of SIET systems and developing efficient algorithms and protocols to improve their performance. Our main interest is in characterizing the fundamental tradeoff between information and energy delivery and the loss incurred by such sampled knowledge. Formally, a sender wishes to send a bitstream of rate $R$, denoted by a uniform message $W \in \{1,\ldots, 2^{nR}\}$. It is encoded into a length $n$ codeword $X^n \in \mathcal{X}^n = [0,1]^n$, which is sent through a memoryless channel $p_{Y|X}$. The interval $[0,1]$ serves as a model for practical discrete-time analog or densely constellated digital communication systems, such as amplitude and frequency modulation (AM and FM), dense constellated quadratic amplitude modulation (QAM) and orthogonal frequency division multiplexing (OFDM) when the input signal space is $\mathcal{X} = [0,1]^2$. For more information, refer to \cite{SeoV2019} for details. When a receiver receives a sequence of signals or waveforms $Y^n \in \mathcal{Y}^n$, it recovers the message $\wh{W}$ corresponding to the bitstream as well as harvests energy. The largest information rate $R$ such that the probability of error $\mathbb{P}[W \ne \wh{W}]$ asymptotically vanishes with $n$ and the harvested energy is greater than $B$ is called the capacity at energy level $B$. The goal is to maximize information and energy delivery simultaneously, i.e., find the optimal tradeoff between information rate and energy delivery. We assume that the information decoder and energy harvester both process the same signal. Our model subsumes a static power-splitting receiver with split ratio $\rho$ \cite{ClerckxZSNKP2019, ZhouZH2012}, up to proper scaling of harvesting function.

The fundamental tradeoff between information rate and energy is characterized by the \emph{capacity-energy} function \cite{Varshney2008}. For a fully known energy harvesting function $b:\mathcal{Y} \mapsto \mathbb{R}_+$, the information capacity for required energy delivery $B$ is $C_b(B)$,
\begin{align*}
	C_b(B) = \max_{p_X: \mathbb{E}[b(Y)] \ge B} I(X;Y),
\end{align*}
where $I(X;Y)$ is the mutual information between the transmitted and received signals \cite{CoverT1991}.
Note that the minimum energy requirement $\mathbb{E}[b(Y)] \ge B$ can be also written in terms of transmission symbol $X$ using conditional expectation,
\begin{align*}
	\mathbb{E}_Y[b(Y)] = \mathbb{E}_X[ \mathbb{E}_{Y|X} [b(Y)] ] = \mathbb{E}_X[\beta(X)],
\end{align*}
where $\beta(x) := \mathbb{E}_{Y|x}[b(Y)]$. This expression is more useful to design transmission symbols or codewords; hence we assume in the sequel that the harvesting function is a function of transmission symbols. It means, with abuse of notation, the capacity-energy function for a new harvesting function $\beta(x)$ is defined as
\begin{align}
	C_{\beta}(B) = \max_{p_X:\mathbb{E}[\beta(X)]\ge B} I(X;Y). \label{def:C_f}
\end{align}

As stated, this work considers a harvesting function that is unknown and accessible only via discrete sample points. Because there are uncountably many candidate functions that yield the same sample points, it is impossible to exactly identify the true function by samples and design optimal SIET systems. Instead, we take a conservative (i.e., worst case) approach that transmits a certain amount of information and energy for \emph{any} harvesting function that agrees on the samples. This leads us to the capacity-energy function for a collection of harvesting functions $F$ agreeing on given samples,
\begin{align}
	C_{F}(B) = \sup_{p_X:\mathbb{E}[\beta(X)]\ge B ~~ \forall \beta \in F} I(X;Y), \label{def:C_F}
\end{align}
which means the maximal information rate at which the energy delivery is no smaller than $B$ for \emph{any} harvesting function in $F$. Clearly, $C_F(B) \le C_\beta(B)$, and $C_\beta(B), C_F(B)$ are non-increasing. Furthermore, $C_{\beta}(B)$ is concave in the interior of the deliverable energy domain,\footnote{The concavity can be readily shown by (coded) time-sharing argument \cite{ElGamalK2011}: For $\alpha > 0$ fraction of time, the system uses a code that achieves $(B_1, C_{\beta}(B_1))$. For the rest of the time, it uses a code that achieves $(B_2, C_{\beta}(B_2))$. Combining these two codes into a single super code achieves the performance $(\alpha B_1 + (1-\alpha)B_2, \alpha C_{\beta}(B_1) + (1-\alpha)C_{\beta}(B_2))$ is achievable, thus there might be a better code achieving higher information rate at $\alpha B_1 + (1-\alpha)B_2$, i.e., the curve is concave.} while $C_{F}(B)$ is not necessarily concave as it is the supremum of concave functions. We also define energy-capacity functions $B_\beta(R), B_F(R)$, dual of the capacity-energy functions,
\begin{align}
	B_{\beta}(R) &= \max_{p_X: I(X;Y) \ge R} \mathbb{E}[\beta(X)], \label{eq:energy_f} \\
	B_{F}(R) &= \max_{p_X: I(X;Y) \ge R} \inf_{\beta \in F} \mathbb{E}[\beta(X)]. \label{eq:energy_F}
\end{align}

Finally, note that according to Shannon's random codebook generation, the capacity-achieving distribution $p_X^*$ that achieves $C_\beta(B)$ (or $C_F(B)$ if $F$ is of interest) can be thought of as an optimal codebook \cite{CoverT1991}.

\subsection{Harvesting Functions and Losses}
Harvesting functions $\beta$ of our interest are of bounded variation.\footnote{Some circuit functions, which may be unstable, fall outside of this category. For instance, if a circuit's nonideality amplifies specific frequency components, it could cause significant fluctuations in frequency, leading to an unbounded variation function.} That is, letting the bound $K > 0$,
\begin{align*}
	\beta \in \BV_K[0,1] := \left\{f: \sup_{P \in \mathcal{P}} \sum_{i=0}^{p-1} |f(x_{i+1})-f(x_i)| \le K \right\},
\end{align*}
where $\mathcal{P}$ is the set of all partitions of $\mathcal{X} = [0,1]$ with $0 \le x_0 \le x_1 \le \cdots \le x_{n_p} = 1$ being partition boundaries. The quantity $\TV(f) = \sup_{P \in \mathcal{P}} \sum_{i=0}^{p-1} |f(x_{i+1})-f(x_i)|$ is called the total variation of $f$. Equivalently, one can write 
\begin{align*}
	\BV_K[0,1] = \left\{f: \int_{\mathcal{X}} |f^{(1)}(x)| \le K \right\},
\end{align*}
with $f^{(1)}$ being the weak derivative of $f$ \cite{AmbrosioFP2000}. Let $\|f\|_{L_p}$ be the $L_p$-functional norm of $f$, i.e.,
\begin{align*}
	\|f\|_{L_p}:= \left( \int |f(x)|^p dx \right)^{1/p}, ~ p \ge 1,
\end{align*}
and $L_2[0,1] = \{f: \|f\|_{L_2} < \infty\}$. Then, two important consequences on $\BV_K[0,1]$ are as follows:
\begin{itemize}
	\item The Jordan decomposition \cite{Billingsley1995} implies that $f \in \BV_K[0,1]$ can be written as a difference of two bounded monotone increasing functions. As each monotone function is in $L_2[0,1]$, we know that $f \in \BV_K[0,1] \subset L_2[0,1]$. 
	
	\item $\BV_K[0,1]$ includes not only Sobolev-smooth harvesting functions considered in \cite{SeoV2019} but also functions with finite number of discontinuities. Hence, the class of our interest is more general than the Sobolev class considered in \cite{SeoV2019}.
\end{itemize}

\begin{figure}[t]
	\centering
	\includegraphics[width=3.5in]{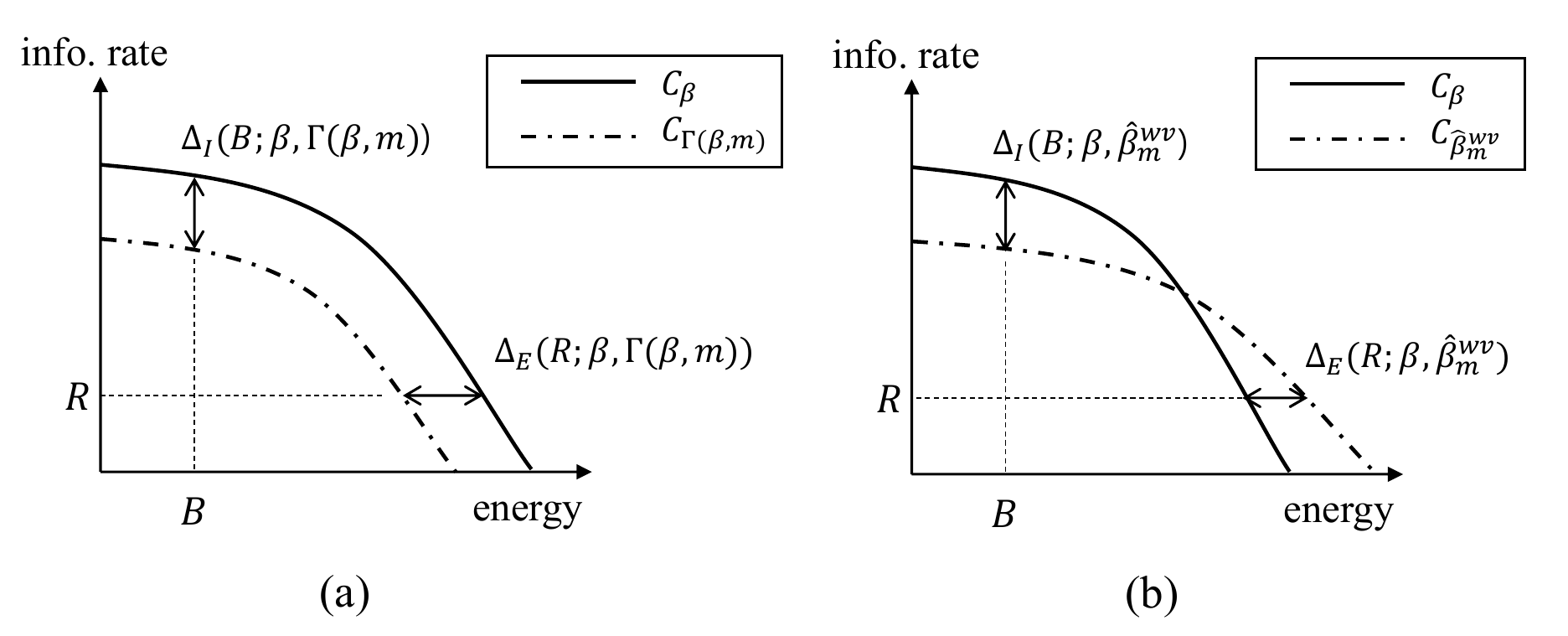}
	\caption{Illustration of typical capacity-energy curves and losses of interest. (a) $C_{\beta}, C_{\Gamma(\beta,m)}$ and induced energy and information losses for noiseless samples. (b) $C_{\beta}, C_{\wh{\beta}_m^{\wv}}$ and losses for a particular realization $\beta_m^{\wv}$ (thus, no expectation here) obtained from noisy samples.}
	\label{fig:def}
\end{figure}

\subsection{Problem Formulation}
A key assumption of this work is that a harvesting function $\beta(x)$ is unknown but available via experimental samples. We consider a \emph{regular grid design} over $\mathcal{X}=[0,1]$ for the location of $m$ sample points, i.e., locations of samples are $\{\tfrac{i}{m-1}: i=0,1,\ldots,m-1 \}$. We consider two settings: The samples are taken in the absence or the presence of noise. The noiseless formulation arises in particular when the transmission symbols' signal-to-noise ratio (SNR) is sufficiently high so that received symbols have no uncertainty and/or harvested energy measurements are reliable, e.g., by taking samples multiple times and averaging them. Then, given samples are $\mathcal{S} = \{ (\tfrac{i}{m-1}, \beta(\tfrac{i}{m-1})) \}_{i=0}^{m-1}$. Having observed $\mathcal{S}$, we can see that there are a number of different but indistinguishable harvesting functions agreeing on the sample points. For an unknown true harvesting function $\beta$, we define $\Gamma(\beta, m) = \{\beta' \in \BV_K[0,1]: \beta'(i/(m-1)) = \beta(i/(m-1)), ~ \forall i \in \{0, \ldots, m-1\} \}$ to be the set of such indistinguishable functions. As we wish to send energy no smaller than $B$ for \emph{any} harvesting function in $\Gamma(\beta, m)$, we consider the worst-case losses in information and energy depicted in Figure \ref{fig:def}(a):
\begin{align}
	\Delta_\ener(R; \beta, \Gamma(\beta,m)) &= B_{\beta}(R) - B_{\Gamma(\beta,m)}(R), \label{def:delta_E} \\
	\Delta_\info(B; \beta, \Gamma(\beta,m)) &= C_{\beta}(B) - C_{\Gamma(\beta,m)}(B). \label{def:delta_I}
\end{align}
Furthermore, the above quantities depend on $\beta$ that we cannot specify; thus knowing them is generally not useful in SIET design. Hence, one may want to guarantee a certain amount of energy for \emph{any} true harvesting functions. To this end, we define the worst-case loss of energy that \emph{does not} depend on unknown $\beta$
\begin{align*}
	\Delta_\ener(R) &= \sup_{\beta \in \BV_K[0,1]} \Delta_\ener(R; \beta, \Gamma(\beta,m)) \\
	&= \sup_{\beta \in \BV_K[0,1]} B_{\beta}(R) - B_{\Gamma(\beta,m)}(R).
\end{align*}
However, we do not take supremum on information loss since the range of deliverable energy is different for each harvesting function, which leads to an arbitrarily large information loss. We will mainly discuss asymptotic bounds on $\Delta_\ener, \Delta_\info$ with the number of samples when we design a SIET system based on the wavelet-reconstructed harvesting function.

Next, assume that samples are taken in the presence of i.i.d.~additive Gaussian noise with mean zero and variance $\sigma^2$. In other words, given noisy samples are 
\begin{align*}
	\mathcal{S} = \{(x_i, Y_i)\}_{i=0}^{m-1} = \{(\tfrac{i}{m-1}, \beta(\tfrac{i}{m-1})+Z_i)\}_{i=0}^{m-1},
\end{align*}
where $Z_i \stackrel{\text{i.i.d.}}{\sim} \mathcal{N}(0, \sigma^2)$. This is the model for low SNR transmission and/or unreliable energy harvesters. Note that once observing noisy samples, we cannot specify a candidate set $\Gamma$ from samples unlike the noiseless case. In other words, $\Gamma = \BV_K$ in this scenario. Since we suggest reconstructing the harvesting function $\wh{\beta}_m^{\wv}$ by wavelets and based on which design a SIET system, we define the expected loss of energy and information as follows. Noting that $\wh{\beta}_m^{\wv}$ is random as samples are random, the expectations are over sampling noise.
\begin{align}
	\overline{\Delta}_\ener(R;\beta, \wh{\beta}_m^{\wv}) &= \mathbb{E}\left[ |B_{\beta}(R) - B_{\wh{\beta}_m^{\wv}}(R)|\right], \label{def:energy_loss_noisy} \\
	\overline{\Delta}_\info(B;\beta, \wh{\beta}_m^{\wv}) &= \mathbb{E}\left[ |C_{\beta}(B) - C_{\wh{\beta}_m^{\wv}}(B)|\right]. \label{def:info_loss_noisy}
\end{align}
Also, note that $\overline{\Delta}_\ener(R;\beta, \wh{\beta}_m^{\wv})$ and $\overline{\Delta}_\info(R;\beta, \wh{\beta}_m^{\wv})$ are dependent on unknown $\beta$. Hence, we ultimately wish to minimize the worst-case loss,
\begin{align*}
	\overline{\Delta}_\ener(R) = \sup_{\beta \in \BV_K[0,1]} \overline{\Delta}_\ener(R;\beta, \wh{\beta}_m^{\wv}).
\end{align*}
However, for the same reason as noiseless samples, we do not take supremum on information loss since the range of deliverable energy is different for each harvesting function, which leads to an arbitrarily large information loss if the supremum is taken.

Lastly, for the problem to be nontrivial, we constraint the input distribution to be (possibly piecewise) continuous and exclude $p_X$ that has point masses. To illustrate the reason, consider a step-increasing harvesting function with a discontinuity at $x_0 \in (0,1)$. Having observed measurement samples, let us suppose our reconstruction is also a step function but with a slightly different location of discontinuity at $x_0' \ne x_0$. Then, some $p_X$ having point masses between $x_0$ and $x_0'$ could show arbitrarily discrepant energy and information rate performance. Discussing conditions on discrete capacity-achieving distributions is beyond our scope; as far as we know, it is an open problem, and the condition for channels to have a discrete capacity-achieving distribution is unknown except for the results on AWGN channels \cite{Smith1971, ShamaiB1995, ChanHK2005, DytsoYPS2020}. We also assume admissible input distributions are bounded, i.e., $p_X(x) \le c_{\maxx}$ for all $x$, because some continuous $p_X$ without bound could play like a distribution having point masses.

\subsection{Wavelets and Multiresolution Analysis}
The wavelet transform is widely used for signal analysis and signal reconstruction from samples because of its excellent empirical performance as well as great theoretical guarantees for a broad range of functions. It represents a function as a sum of expanding wave-like basis functions $\{\psi_{j,k}\}_{j,k}$ constructed from a scaling function (or father wavelet) $\phi(x)$. We only provide a minimal introduction here; see \cite{Daubechies1992, Mallat2008} for further details.

Let $V_j$ be the space at scale $j$ that can be spanned by a linear combination of $\{2^{-j/2}\phi(2^j x-k)\}_{k \in \mathbb{Z}}$. It is called a multiresolution structure if the resulting $V_j$'s satisfy
\begin{align*}
	\cdots \subset V_{-2} \subset V_{-1} \subset V_0 \subset V_1 \subset V_2 \subset \cdots
\end{align*}
with $\cap_{j} V_{j} = \{0\}$ and $\cup_j V_{j}$ being dense in $L_2(\mathbb{R})$. Larger scales mean finer resolution. Moreover, $V_{j+1} = V_{j} \oplus V_j^{\perp}$ where $\oplus$ is the direct sum and $V_j^{\perp}$ is the subspace of functions in $V_{j+1}$ that are orthogonal to $V_j$. Then, we can define a \emph{mother} wavelet from $\phi$ such that its linear combinations of scaling and translation form a basis for $W_0$. Letting $\psi$ be the mother wavelet, we can define a child wavelet at scale $j$ and location $k$ as $\psi_{j,k}(x) = 2^{-j/2} \psi(2^jx-k)$. A notable property of child wavelets is that $\psi_{j,k}$ are orthogonal to $\psi_{j',k'}$ unless $(j,k) = (j',k')$.

\begin{figure}[t]
	\centering
	\includegraphics[width=3.0in]{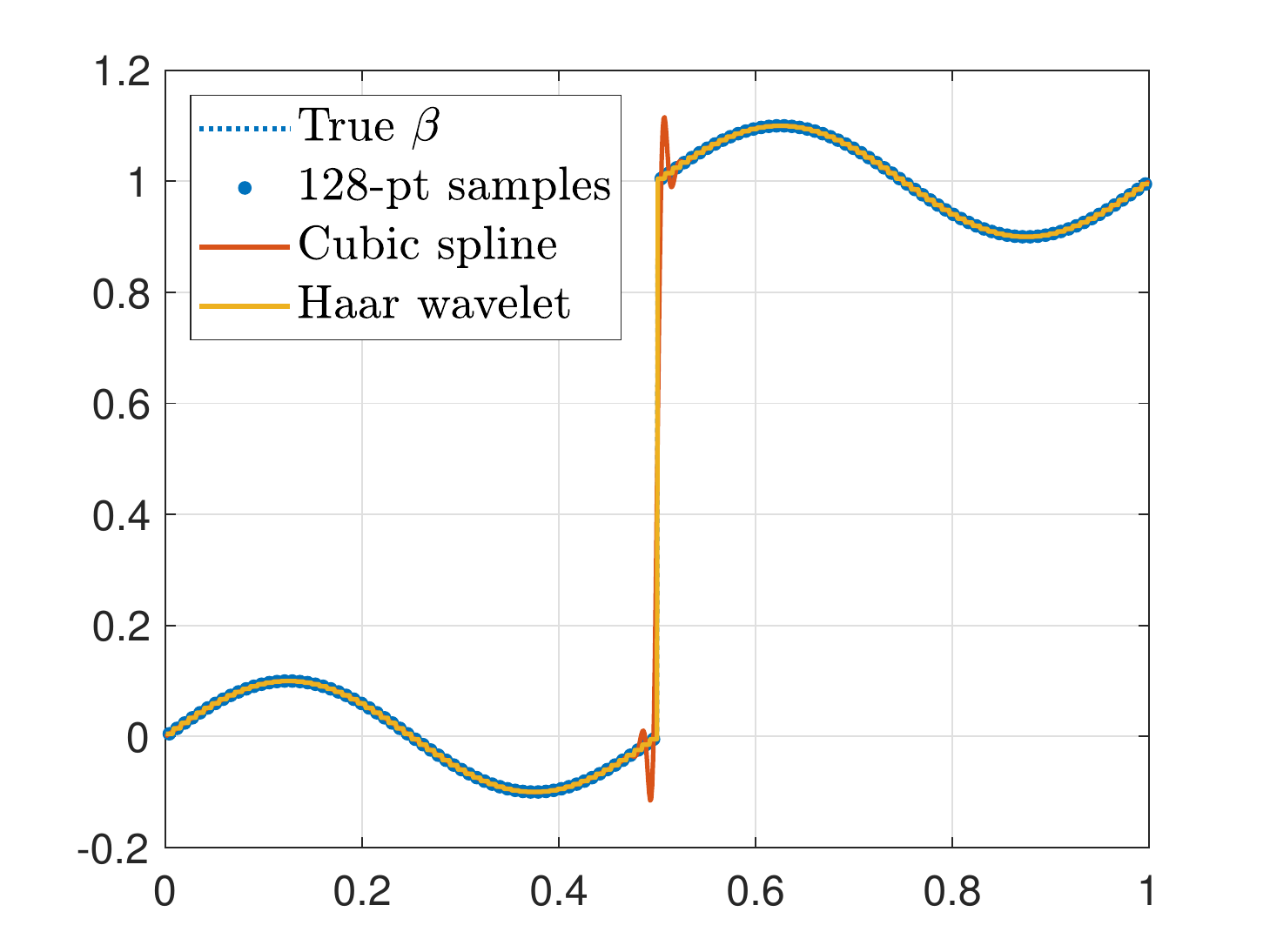}
	\caption{Example of wavelet and spline reconstruction for a discontinuous function. The true function (dotted) is $f(x) = 0.1\sin(4\pi x) + \bm{1}_{ \{ x \ge 0.5\} }$, and $128$-point samples are taken. The cubic spline (red) manifests under- and overshoots near the discontinuity, while the Haar wavelet (yellow) tracks the discontinuity closely. }
	\label{fig:sp_vs_wv}
\end{figure}

The simplest wavelet is perhaps the Haar wavelets, where $\phi(x) = \bm{1}_{\{0 \le x \le 1\}}(x)$ and $\psi(x) = \phi(2x) - \phi(2x-1)$. Another popular one is Daubechies wavelets that enjoy the property of ``$p$-vanishing moment'' on a finite-length support, see \cite{Daubechies1992, Mallat2008}. An interesting class of wavelets is \emph{spline wavelets} \cite{Unser1997}, the bases of which are generated from splines. Wavelets offer comparable reconstruction performance to that of non-wavelet methods (e.g., splines \cite{Unser1999}) for some good classes of functions, e.g., Sobolev space, and in general offer better performance than non-wavelet ones for a broader class of functions. Figure \ref{fig:sp_vs_wv} depicts its reconstruction accuracy for a discontinuous function compared to that of the spline method. Even with a sufficient number of samples ($128$ points), the spline-reconstructed function reveals fundamental inaccuracy near discontinuities.

\section{Noiseless Samples} \label{sec:noiseless_samples}
This section discusses the asymptotic characterization of $\Delta_\ener, \Delta_\info$ with the number of samples $m$ when samples are taken without measurement noise. Note that definitions \eqref{def:delta_E} and \eqref{def:delta_I} only characterize the performance loss; bounds on the quantities alone do not immediately suggest how to achieve them. To that end, we suggest using wavelet reconstruction for a harvesting function and designing a SIET system as if the wavelet-reconstructed harvesting function is the true one. As we will see, it will attain asymptotically negligible performance loss with the number of samples.

Before proceeding, the following proposition characterizes the function reconstruction loss incurred by wavelets.
\begin{prop} \label{prop:noiseless}
	Let $f \in \BV_K[0,1]$ be the function to be reconstructed, and $m=2^J$ observations $\{( i/(m-1), f(i / (m-1)) ) \}_{i=0}^{2^J-1}$ are given. Then, there exists a wavelet reconstruction from observations at scale $J$, $\wh{f}_J^{\wv} \in V_J$, such that
	\begin{align*}
		\|f-\wh{f}_J^{\wv}\|_{L_2}^2 \le c \TV^2(f) m^{-1} = \mathcal{O}(m^{-1}).
	\end{align*}
	with some positive constant $c$.
\end{prop}
\begin{IEEEproof}
	The proof can be found in Appendix \ref{app:pf_prop1}.
\end{IEEEproof}
This proposition will be used as a building block in the next subsections.

\subsection{Energy Loss}
The following theorem is our main result on energy loss. The first statement shows that the energy loss $\Delta_\ener(R)$ is asymptotically negligible at all information rates. However, the claim alone does not give us a constructive answer on how to design a SIET system that achieves such performance since $\Delta_\ener(R)$ requires optimization over all functions in $\Gamma(\beta,m)$. The second statement shows how to design a codebook---na\"{i}vely believing $\wh{\beta}_m^{\wv}$ is the truth and designing a codebook based on it achieve the asymptotics.
\begin{thm} \label{thm:delta_E_noiseless}
	It holds that
	\begin{align*}
		\Delta_\ener(R) = \mathcal{O}(m^{-1/2}) ~~ \forall R \ge 0.
	\end{align*}
	Moreover, designing an optimal codebook as if the wavelet-reconstructed harvesting function $\wh{\beta}_m^{\wv}$ is the true harvesting function achieves $\mathcal{O}(m^{-1/2})$ energy loss.
\end{thm}
\begin{IEEEproof}
	Fix an arbitrary (possibly piecewise) continuous input distribution $p_X$ that is bounded on $[0,1]$, i.e., $p_X(x) \le c_{\maxx}$ for all $x \in \mathcal{X}$. Also, consider a random codebook generated from $p_X$. Let us pick an arbitrary $\wh{\beta}_m(X) \in \Gamma(\beta, m) \subset \BV_K[0,1]$ that is not necessarily wavelet-reconstructed. Then, 
	\begin{align*}
		&\left( \mathbb{E}_{p_X}[\beta(X)] - \mathbb{E}_{p_X}[\wh{\beta}_m(X)] \right)^2 = \left( \mathbb{E}_{p_X}[\beta(X) - \wh{\beta}_m(X)] \right)^2 \\
		&\stackrel{(a)}{\le} \mathbb{E}_{p_X} \left[ (\beta(X) - \wh{\beta}_m(X) )^2 \right] = \int_{\mathcal{X}} p_X(x) (\beta(x) - \wh{\beta}_m(x))^2 dx \\
		&\le c_{\maxx} \int_{\mathcal{X}} (\beta(x) - \wh{\beta}_m(x))^2 dx = c_{\maxx} \|\beta - \wh{\beta}_m\|_{L_2}^2,
	\end{align*}
	where (a) follows from Jensen's inequality. Taking the square root on both sides and telescoping with wavelet-reconstructed harvesting function $\wh{\beta}^{\wv}$,
	\begin{align}
		&\left| \mathbb{E}_{p_X}[\beta(X)] - \mathbb{E}_{p_X}[\wh{\beta}_m(X)] \right| \le \sqrt{c_{\maxx}}\|\beta - \wh{\beta}_m\|_{L_2} \nonumber \\
		&\le \sqrt{c_{\maxx}}\|\beta - \wh{\beta}_m^{\wv}\|_{L_2} + \sqrt{c_{\maxx}}\|\wh{\beta}_m^{\wv} - \wh{\beta}_m\|_{L_2} \nonumber \\
		&= \mathcal{O}(m^{-1/2}) + \mathcal{O}(m^{-1/2}) = \mathcal{O}(m^{-1/2}), \label{eq:energy_loss}
	\end{align}
	where Proposition \ref{prop:noiseless} is used twice since $\beta, \wh{\beta}_m \in \BV_K[0,1]$.
	
	Note that the above result holds for any $p_X, \beta, \wh{\beta}_m \in \Gamma(\beta, m)$. Let $p_X^*, q_X^*$ be the capacity-achieving distributions of $B_{\beta}, B_{\Gamma(\beta,m)}$ at rate $R$, respectively. That is, $B_{\beta} = \mathbb{E}_{p_X^*}[\beta(X)]$ and $B_{\Gamma(\beta,m)} = \min_{\wh{\beta}_m \in \Gamma(\beta,m)} \mathbb{E}_{q_X^*}[\wh{\beta}(X)]$. Then,
	\begin{align*}
		B_{\beta}(R) &\stackrel{(b)}{\ge} B_{\Gamma(\beta,m)}(R) = \min_{\wh{\beta}_m \in \Gamma(\beta,m)} \mathbb{E}_{q_X^*}[\wh{\beta}_m(X)] \\
		&\stackrel{(c)}{\ge} \min_{\wh{\beta}_m \in \Gamma(\beta,m)} \mathbb{E}_{p_X^*}[\wh{\beta}_m(X)] \\
		&\stackrel{(d)}{\ge} \mathbb{E}_{p_X^*}[\beta(X)] - c m^{-1/2} \\
		&= B_{\beta}(R) - c m^{-1/2},
	\end{align*}
	where (b) follows from the definitions \eqref{eq:energy_f} and \eqref{eq:energy_F}, (c) follows since $p_X^*$ is suboptimal for $B_{\Gamma(\beta,m)}(R)$, and (d) follows from \eqref{eq:energy_loss}. The result implies that $\Delta_{\ener}(R; \beta, \Gamma(\beta, m)) = \mathcal{O}(m^{-1/2})$ for all $\beta \in \BV_K[0,1]$ since $\beta$ is arbitrary. Taking supremum over $\beta \in \BV_K[0,1]$ concludes the first claim.
	
	To show the second claim, fix $R \ge 0$ and consider a harvesting function $\beta$ and its wavelet reconstruction $\wh{\beta}_m^{\wv}$. As $\wh{\beta}_m^{\wv} \in \Gamma(\beta, m)$, we have 
	\begin{align}
		B_{\beta}(R)-B_{\wh{\beta}_m^{\wv}}(R) \le B_{\beta}(R) - B_{\Gamma(\beta,m)}(R) \le cm^{-1/2}. \label{eq:temp1}
	\end{align}
	Note that if $\wh{\beta}_m^{\wv}$ is the true one, then $\beta \in \Gamma(\wh{\beta}_m^{\wv}, m)$ since $\beta, \wh{\beta}_m^{\wv}$ both agree on the sample points. This implies 
	\begin{align}
		B_{\wh{\beta}_m^{\wv}}(R)-B_{\beta}(R) \le cm^{-1/2}. \label{eq:temp2}
	\end{align}
	
	Assume that capacity-achieving distributions for $B_{\beta}(R)$, $B_{\wh{\beta}_m^{\wv}}(R)$ are $p_X^*, q_X^*$ respectively, based on which the optimal codebooks are generated. Then, $q_X^*$-generated codebook for $\wh{\beta}_m^{\wv}$ will deliver energy $\mathbb{E}_{q_X^*}[\beta(X)]$ under $\beta$ and $B_{\beta}(R) = \mathbb{E}_{p_X^*}[\beta(X)] \ge \mathbb{E}_{q_X^*}[\beta(X)]$. Therefore,
	\begin{align*}
		&|B_{\beta}(R) - \mathbb{E}_{q_X^*}[\beta(X)]| \\
		&\stackrel{(e)}{\le} |B_{\beta}(R) - B_{\wh{\beta}_m^{\wv}}(R)| + |B_{\wh{\beta}_m^{\wv}}(R) - \mathbb{E}_{q_X^*}[\beta(X)]| \\
		&\stackrel{(f)}{\le} cm^{-1/2} + |B_{\wh{\beta}_m^{\wv}}(R) - \mathbb{E}_{q_X^*}[\beta(X)]| \\
		&= cm^{-1/2} + \left| \mathbb{E}_{q_X^*}[\wh{\beta}_m^{\wv}(X)] - \mathbb{E}_{q_X^*}[\beta(X)] \right| \\
		&\stackrel{(g)}{\le} cm^{-1/2} + cm^{-1/2} = \mathcal{O}(m^{-1/2}),
	\end{align*}
	where (e) follows from the triangle inequality, (f) from \eqref{eq:temp1} and \eqref{eq:temp2}, and (g) from \eqref{eq:energy_loss}. Hence, we can conclude that by designing a codebook based on $\wh{\beta}_m^{\wv}$, we can deliver energy with at most $\mathcal{O}(m^{-1/2})$ loss from the best operating point.
\end{IEEEproof}

The above theorem discusses the upper bound on the energy loss $\mathcal{O}(m^{-1/2})$. Hence one may wonder whether the bound is asymptotically tight or not. In function approximation theory, if obtained (wavelet) coefficients are not further processed, it is called \emph{linear} approximation since obtaining coefficients is nothing but a linear projection. Interestingly, it is known that the bound in Proposition \ref{prop:noiseless} is the best among all linear approximation methods \cite{Mallat2008, deVoreL1993, DeVore1998}. Hence, our bound $\mathcal{O}(m^{-1/2})$ is also the best among all linear approximation methods. To outperform $\mathcal{O}(m^{-1/2})$, \emph{nonlinear} approximation should be adopted, i.e., obtained wavelet coefficients need further processing. This is because if a target function has a discontinuity, wavelet coefficients related to locations close to the discontinuity are larger, thus more important. However, since the discontinuity's location is unknown, samples should be taken at $m' > m$ equidistant points, and only $m$ largest (i.e., important) ones should be selected (i.e., processed), or equivalently $m$ sampling points should be placed adaptively near the discontinuity. This method, belonging to nonlinear approximation, is known to achieve $\|f-\wh{f}_m^{\wv}\|_{L_2}^2 = \Theta(m^{-2})$ for $\BV_K[0,1]$ space \cite{Mallat2008, deVoreL1993, DeVore1998}, i.e., no other methods improve the rate $m^{-2}$. However, such a nonlinear $m$-term approximation is not applicable in our application because sampling locations should be predetermined before measurements. Hence, it is beyond our interest.

Having said that $\mathcal{O}(m^{-1/2})$ is tight among methods without further coefficient manipulation, it is interesting to compare it to the decay rate for Sobolev classes: Theorems 4 and 7 in \cite{SeoV2019} state that the energy loss for the functions in Sobolev class with differentiability parameter $\lambda \ge 1$ is $\mathcal{O}(m^{-\lambda})$, which is tight for some functions. As $\BV_K[0,1]$ includes more functions than the Sobolev class, the difference $\lambda - \frac{1}{2}$ in the exponent tells that the extra functions such as discontinuous functions impose the penalty $\lambda - \frac{1}{2}$ in the exponent. If our harvesting functions are restricted to the Sobolev class with parameter $\lambda$, the wavelet reconstruction still recovers $\mathcal{O}(m^{-\lambda})$. That is, the wavelet-based transmission is also optimal. The proof immediately follows from \cite[Chapter 9]{Mallat2008} and repeating the above argument, so omitted.

\subsection{Information Loss}
Moving our focus to the information-rate loss, we show in the following theorem and corollary that the information loss is also asymptotically negligible. However, the difference is that, unlike the energy loss that asymptotically vanishes at all $R \ge 0$, the information loss vanishes only on the interior of the deliverable energy range. One should be cautious when designing a SIET system at maximal energy delivery since the information loss could be arbitrary.
\begin{thm} \label{thm:delta_I_noiseless}
	For any $\beta \in \BV_K[0,1]$ and $B \in [0, B_{\maxx})$,
	\begin{align*}
		\Delta_\info(B; \beta, \Gamma(\beta,m)) \to 0 \textrm{ as } m \to \infty.
	\end{align*}
	However, assuming the existence of a channel such that the capacity-achieving distribution is nonvanishing on $\mathcal{X}$, i.e., for all $x$, $p_X(x) > c$ with some $c>0$, $\Delta_{\textsf{I}}(B_{\maxx};\beta, \Gamma(\beta,m)) = C_{\maxx}$ for some $\beta \in \BV_K[0,1]$, where $C_\maxx$ is the unconstrained capacity i.e., $C_\maxx = \max_{p_X} I(X;Y)$.
\end{thm}
\begin{IEEEproof}
	First, at $B=0$, the capacity-energy function is indeed the same as unconstrained capacity. In other words, any input probability is admissible, and then $C_{\beta}(0)=C_{\wh{\beta}_m}(0)=C_\maxx$. Hence, $\Delta_\info(0; \beta, \Gamma(\beta,m)) = 0$.
	
	Consider $B \in (0, B_\maxx)$. Recall that $C_{\beta}(B)$ is concave, implying that $C_{\beta}(B)$ is continuous on the interior of the energy domain. Then, Theorem \ref{thm:delta_E_noiseless} implies that for any $B$, there exists $B'$ such that $C_{\Gamma(\beta,m)}(B) = C_{\beta}(B')$ and $|B-B'| = \mathcal{O}(m^{-1/2})$. That is,
	\begin{align*}
		&\Delta_\info(B;\beta, \Gamma(\beta, m)) = C_{\beta}(B) - C_{\Gamma(\beta, m)}(B) \\
		&= C_{\beta}(B) - C_{\beta}(B') \\
		&= C_{\beta}(B) - C_{\beta}(B + \mathcal{O}(m^{-1/2})).
	\end{align*}
	As $C_{\beta}(R)$ is continuous, $\Delta_\info(B;\beta, \Gamma(\beta, m)) \to 0$ as the number of samples grows. This proves the first claim.
	
	To show the second claim, assuming the existence of a stated channel, consider a constant harvesting function, say $\beta(x) = M$. Obviously, any $p_X$ is admissible if $B \le M$ and none is if $B > M$. Therefore,
	\begin{align*}
		C_{\beta}(B) = \begin{cases}
			C_{\maxx} & \text{ if } B \le M = B_{\maxx}, \\
			0 & \text{ if } B > M = B_{\maxx}.
		\end{cases}
	\end{align*}
	
	Note that $\Gamma(\beta, m)$ includes some $\beta'$ such that $\beta'(x) = \beta(x) = M$ at sample points, but $\beta'(x) < \beta(x)$ otherwise, cf. a nonnegative bumpy function idea \cite[the proof of Theorem 7]{SeoV2019} on this. Hence, $\mathbb{E}[\beta'(X)] < M$ for $p_X$ such that $p_X(x) > c$, which in turn implies $C_{\Gamma(\beta,m)}(M) = 0$. In other words, it is impossible to transfer energy $M = B_{\maxx}$ for all harvesting functions in $\Gamma(\beta, m)$. It is an example of the second claim such that $\Delta_{\textsf{I}}(B_{\maxx};\beta, \Gamma(\beta,m)) = C_{\maxx}$.
\end{IEEEproof}

The above theorem only discusses the convergence, not the speed of convergence with $m$. Analytically characterizing the speed of convergence of the information loss is generally challenging because it highly depends on the channel and its capacity-achieving distribution. The following corollary circumvents such difficulty by assuming $C_{\beta}(B)$ is $\alpha$-H\"{o}lder continuous, which is a generalized version of the Lipschitz continuity.
\begin{cor} \label{cor:holder_noiseless}
	Suppose that for a harvesting function $\beta \in \BV_K[0,1]$ and a channel, the energy-capacity function is $\alpha$-H\"{o}lder continuous. That is, for any $B_1, B_2 \in (0, B_{\maxx})$,
	\begin{align*}
		|C_{\beta}(B_1) - C_{\beta}(B_2)| \le c |B_1 - B_2|^{\alpha}
	\end{align*}
	with some constant $c > 0$. Then, $\Delta_\info(B; \beta, \Gamma(\beta,m)) = \mathcal{O}(m^{-\alpha/2})$ for $B \in [0, B_{\maxx})$.
\end{cor}
\begin{IEEEproof}
	First, consider the case when $B=0$. As it reduces to the unconstrained capacity, any $p_X$ is admissible. Hence, $C_{\beta}(0) = C_{\Gamma(\beta,m)}(0)$ for any $\beta$, and $\Delta_\info(B; \beta, \Gamma(\beta,m))=0$.
	
	For $B \in (0, B_{\maxx})$, 
	\begin{align*}
		&\Delta_\info(B; \beta, \Gamma(\beta,m)) = C_{\beta}(B) - C_{\Gamma(\beta,m)}(B) \\
		&\stackrel{(a)}{=} C_{\beta}(B) - C_{\beta}(B+\mathcal{O}(m^{-1/2})) \\
		&\stackrel{(b)}{\le} \text{const} \cdot m^{-\alpha/2} = \mathcal{O}(m^{-\alpha/2}),
	\end{align*}
	where (a) follows from Theorem \ref{thm:delta_E_noiseless} and (b) follows since $C_{\beta}$ is $\alpha$-H\"{o}lder. The claim is proved.
\end{IEEEproof}
Note that when $\alpha=1$, i.e., when $C_{\beta}(B)$ is Lipschitz continuous, the information loss reduces to $\Delta_\info(B; \beta, \Gamma(\beta, m)) = \mathcal{O} \left( m^{-1/2} \right)$, which is the same as the energy loss asymptotics.

\section{Noisy Samples} \label{sec:noisy_samples}
This section discusses the expected losses of energy and information when measurement samples are noisy. As only noisy samples are available, we suggest reconstructing a harvesting function by wavelets with coefficients being soft-thresholded for denoising and then designing a SIET system as if the final reconstructed harvesting function is the true one. It turns out that the loss of energy and information incurred by this is asymptotically negligible and tight up to a logarithmic factor.

The proposition below characterizes the function reconstruction loss incurred by wavelets when samples are noisy. Unlike Proposition \ref{prop:noiseless} that uses linear approximation from noiseless samples, the next proposition is by nonlinear approximation, i.e., it processes coefficients. It first approximates $f \in \BV_K[0,1]$ by $m$-coefficient wavelets where obtained coefficients are noisy as well, and then manipulating coefficients by soft thresholding \cite{DonohoJ1994}. In some literature, the latter manipulation is referred to as \emph{wavelet shrinkage} since soft thresholding ``shrinks'' the magnitude of coefficients. It is known to be universal and nearly minimax optimal in the sense that for most function classes such as Sobolev, Besov, H\"{o}lder, and bounded variation function classes, its reconstruction performance is minimax optimal up to the logarithmic prefactor \cite{Donoho1995}. For non-wavelet reconstruction methods that are optimal for some smooth function classes, see \cite[Chapters 1, 2]{Tsybakov2009}.

\begin{prop} \label{prop:noisy_wavelet_error}
	Let $f \in \BV_K[0,1]$ be the function to be recovered, and $m=2^J$ noisy observations $\mathcal{S} = \{(x_i, Y_i)\}_{i=0}^{2^J-1} = \{(i/2^J, f(i / 2^{J}) + Z_i) \}_{i=0}^{2^J-1}$ are given. Assume $Z_i$'s are independent and identically distributed (i.i.d.) Gaussian with mean zero and variance $\sigma^2$. Then, wavelet reconstruction $\wh{f}^{\wv}$ by soft-thresholding estimation of coefficients attains the expected error
	\begin{align*}
		\mathbb{E}_Z \left[ \|f-\wh{f}^{\wv}\|_{L_2}^2 \right] = \mathcal{O}(m^{-2/3} \log m).
	\end{align*}
\end{prop}
\begin{IEEEproof}
	The proof can be found in Appendix \ref{app:pf_prop2}.
\end{IEEEproof}
Like the case of noiseless samples, the above proposition will be used as a building block in the next subsections.

\subsection{Energy Loss}	
The next theorem is on the expected energy loss, which shows that if the above wavelet reconstruction is used, the loss of energy incurred by the partial knowledge asymptotically vanishes.
\begin{thm} \label{thm:delta_E_noisy}
	$\overline{\Delta}_\ener(R) = \mathcal{O}(m^{-1/3} \sqrt{\log m}) ~~ \forall R \ge 0$. Moreover, designing an optimal codebook as if the wavelet-reconstructed harvesting function $\wh{\beta}_m^{\wv}$ is the true harvesting function achieves the asymptotics.
\end{thm}
\begin{IEEEproof}
	Note that by Jensen's inequality,
	\begin{align*}
		\left( \mathbb{E}_Z\left[ \|\beta - \wh{\beta}_m^{\wv}\|_{L_2} \right] \right)^2 &\le \mathbb{E}_Z \left[ \|\beta - \wh{\beta}_m^{\wv}\|_{L_2}^2 \right] \\
		&= \mathcal{O}(m^{-2/3} \log m).
	\end{align*}
	Therefore, we have
	\begin{align}
		\mathbb{E}_Z \left[ \|\beta - \wh{\beta}_m^{\wv}\|_{L_2} \right] &= \mathcal{O}(m^{-1/3} \sqrt{\log m}). \label{eq:after_jensen}
	\end{align}
	
	Let $p_X^*, q_X^*$ be the capacity-achieving distributions of $B_{\beta}(R), B_{\wh{\beta}_m^{\wv}}(R)$, respectively. Recall that distributions of our interest are (possibly piecewise) continuous and bounded on $[0,1]$, i.e., $p_X^*(x), q_X^*(x) \le c_{\maxx}$ for all $x \in \mathcal{X}$. Also, recall that a distribution generates a random codebook. Let $E$ be the set of noise events such that $B_{\beta}(R) \le B_{\wh{\beta}_m^{\wv}}(R)$. Then,
	\begin{align*}
		&\overline{\Delta}_\ener(R;\beta, \wh{\beta}_m^{\wv}) = \mathbb{E}_Z\left[ |B_{\beta}(R) - B_{\wh{\beta}_m^{\wv}}(R)|\right] \\
		&= \mathbb{E}_Z\left[ (B_{\wh{\beta}_m^{\wv}}(R) - B_{\beta}(R))\bm{1}_{\{ E \}} + (B_{\beta}(R) - B_{\wh{\beta}_m^{\wv}}(R))\bm{1}_{\{ E^c \}} \right].
	\end{align*}
	
	Consider the first term corresponding to the event $E$.
	\begin{align*}
		&B_{\wh{\beta}_m^{\wv}}(R) - B_{\beta}(R) \\
		&= \mathbb{E}_{q_X^*} [\wh{\beta}_m^{\wv}(X)] - \mathbb{E}_{p_X^*} [\beta(X)] \\
		&\stackrel{(a)}{\le} \mathbb{E}_{q_X^*} [\wh{\beta}_m^{\wv}(X)] - \mathbb{E}_{q_X^*} [\beta(X)] \\
		&\le \int_{\mathcal{X}} q_X^*(x) |\wh{\beta}_m^{\wv}(x) - \beta(x)| dx \\
		&\stackrel{(b)}{\le} c_{\maxx} \int_{\mathcal{X}} |\beta(x) - \wh{\beta}_m^{\wv}(x) | dx \\
		&= c_{\maxx} \| \beta - \wh{\beta}_m^{\wv} \|_{L_1},
	\end{align*}
	where (a) follows since $q_X^*$ is suboptimal for $\beta$, replacing $p_X^*$ with $q_X^*$ increases the loss, and (b) follows since $q_X^*$ is bounded. Using \eqref{eq:after_jensen} and the $L_p$-norm monotonicity for a finite measure space, we have the following bound on the first term.
	\begin{align*}
		&\mathbb{E}_Z\left[ (B_{\wh{\beta}_m^{\wv}}(R) - B_{\beta}(R))\bm{1}_{\{ E \}} \right] \\
		&\le c_\maxx \mathbb{E}_Z \left[ \|\beta - \wh{\beta}_m^{\wv}\|_{L_1} \bm{1}_{\{ E \}} \right] \\
		&\le c_\maxx \mathbb{E}_Z \left[ \|\beta - \wh{\beta}_m^{\wv}\|_{L_2} \bm{1}_{\{ E \}} \right] \\
		&\le c_\maxx \mathbb{E}_Z \left[ \|\beta - \wh{\beta}_m^{\wv}\|_{L_2} \right] =\mathcal{O}(m^{-1/3} \sqrt{\log m}).
	\end{align*}
	
	Similarly, for the event $E^c$, we can obtain the same bound $\mathcal{O}(m^{-1/3} \sqrt{\log m})$, thus, $\overline{\Delta}_\ener(R;\beta, \wh{\beta}_m^{\wv}) = \mathcal{O}(m^{-1/3} \sqrt{\log m})$.
	
	Finally, note that the asymptotic bound does not depend on $\beta$ as long as $\beta \in \BV_K[0,1]$. Hence, taking supremum over $\beta \in \BV_K[0,1]$ does not change the bound, i.e.,
	\begin{align*}
		\overline{\Delta}_\ener(R) = \sup_{\beta \in \BV_K[0,1]} \overline{\Delta}_\ener(R;\beta, \wh{\beta}_m^{\wv}) = \mathcal{O}(m^{-1/3} \sqrt{\log m}).
	\end{align*}
	
	To show the second claim, we can repeat the above argument and the proof of Theorem \ref{thm:delta_E_noiseless} as follows.
	\begin{align*}
		&B_{\beta}(R) - \mathbb{E}_{q_X^*}[\beta(X)] = \left| B_{\beta}(R) - \mathbb{E}_{q_X^*}[\beta(X)] \right| \\
		&\le \left| B_{\beta}(R) - B_{\wh{\beta}_m^{\wv}}(R) \right| + \left| B_{\wh{\beta}_m^{\wv}}(R) - \mathbb{E}_{q_X^*}[\beta(X)] \right| \\
		&= \left| B_{\beta}(R) - B_{\wh{\beta}_m^{\wv}}(R) \right| \bm{1}_{\{ E \}}  + \left| B_{\beta}(R) - B_{\wh{\beta}_m^{\wv}}(R) \right| \bm{1}_{\{ E^c \}} \\
		&~~~~~~~~~~~~~~~~~~~~~~~~~~~~ + \left| B_{\wh{\beta}_m^{\wv}}(R) - \mathbb{E}_{q_X^*}[\beta(X)] \right| \\
		&\le c_{\maxx} \| \beta - \wh{\beta}_m^{\wv} \|_{L_1} (\bm{1}_{\{ E \}} + \bm{1}_{\{ E^c \}}) \\
		&~~~~~~~~~~~~~~~~~~~~~~~~~~~~ + \left| \int q_X^*(x) (\wh{\beta}_m^{\wv}(x) - \beta(x)) dx \right| \\
		&\le c_{\maxx} \| \beta - \wh{\beta}_m^{\wv} \|_{L_1} + c_{\maxx} \int |\wh{\beta}_m^{\wv}(x) - \beta(x)| dx \\
		&= 2 c_{\maxx} \| \beta - \wh{\beta}_m^{\wv} \|_{L_1}.
	\end{align*}
	Taking expectation over noise, applying the $L_p$-norm monotonicity, and \eqref{eq:after_jensen} yield
	\begin{align*}
		&\mathbb{E}_Z \left[ B_{\beta}(R) - \mathbb{E}_{q_X^*}[\beta(X)] \right] \le \mathbb{E}_Z [2 c_{\maxx} \| \beta - \wh{\beta}_m^{\wv} \|_{L_1} ] \\
		&\le 2 c_{\maxx} \mathbb{E}_Z [ \| \beta - \wh{\beta}_m^{\wv} \|_{L_2} ] = \mathcal{O}(m^{-1/3} \sqrt{\log m}).
	\end{align*}
	It proves the second claim.
\end{IEEEproof}

As the wavelet reconstruction is a specific instance of various reconstruction methods, it is natural to ask whether or not the rate $\mathcal{O}(m^{-1/3} \sqrt{\log m})$ is asymptotically optimal. We have an affirmative answer that the $\wh{\beta}_m^{\wv}$-based SIET design is asymptotically optimal for energy delivery up to a logarithmic factor, i.e., in the sense of $\widetilde{\mathcal{O}}(\cdot)$.\footnote{$\widetilde{\mathcal{O}}(\cdot), \widetilde{\Theta}(\cdot)$ ignore a logarithmic factor, for instance, $\widetilde{\mathcal{O}}(m^{-1/3}) = \widetilde{\mathcal{O}}(m^{-1/3} \log^k m)$ for any $k \in \mathbb{N}$.} The next theorem indeed proves $\overline{\Delta}_\ener(R) = \Omega(m^{-1/3})$ in a hypothetical setting. Therefore, $\overline{\Delta}_\ener(R) = \widetilde{\Theta}(m^{-1/3})$.

\begin{thm} \label{thm:E_lowerbound_noisy}
	For some channel and rate $R$, $\overline{\Delta}_\ener(R) = \Omega(m^{-1/3})$.
\end{thm}
\begin{IEEEproof}
	Assume a hypothetical channel such that its unique unconstrained capacity-achieving distribution $p_X^*$ is a square pulse: $p_X^*(x)= c > 0$ on every interval $I_i:=[i/r, (i+0.5)/r]$ for $i \in \{0,1,\ldots, r-1\}$, and $p_X^*(x)= 0$ otherwise. The number of bins $r$ will be specified later.
	
	Instead of $\BV_K[0,1]$ that has uncountably many functions, we assume that the harvesting function is in a finite subset $F \subset \BV_K[0,1]$. Let us define
	\begin{align*}
		F := \left\{ \beta_{\mathbf{b}}(x): \beta_{\mathbf{b}}(x) = A\left(1+\sum_{i=0}^{r-1} b_i \psi(rx-i)\right) \right\},
	\end{align*}
	where $b_i$'s are $\pm 1$ and $\psi$ is the Haar wavelet. Therefore, each binary sequence is mapped to a distinct piecewise constant harvesting function, and there are $2^r$ functions in $F$ in total. As the Haar wavelet has two discontinuities in each interval of length $1/r$, the total variation of $\beta_{\mathbf{b}}$ is at most $4Ar$. Let $A = \frac{K}{4r}$ so that $F \subset \BV_K[0,1]$. Now, we will detect $\beta_{\mathbf{b}}$ accurately from noisy samples, or equivalently detect all $b_i$'s correctly, and based on which design SIET. Hence, the original function estimation problem is now turned into a lower bound on a detection problem with finite hypotheses, which is easier.
	
	Note that if $b_i = 1$, the samples are distributed according to $\mathcal{N}(2A, \sigma^2)$ on the first half of the interval and $\mathcal{N}(0, \sigma^2)$ on the second half. If $b_i = -1$, it is the opposite. Since we have $m$ equidistant samples and each Haar wavelet is supported on a disjoint interval of length $1/r$, there are $m/r$ samples in each interval. Therefore, the problem of detecting each $b_i$ is indeed a binary hypothesis testing problem between two Gaussians:
	\begin{align*}
		H_0: Y \sim \mathcal{N}(2A, \frac{r\sigma^2}{m}) ~~ \text{and} ~~ H_1: Y \sim \mathcal{N}(0, \frac{r\sigma^2}{m}).
	\end{align*}
	As the likelihood ratio test is optimal, it produces the smallest error probability $Q(\frac{A \sqrt{m}}{\sqrt{r}\sigma})$, where $Q(\cdot)$ is the $Q$-function. Using $A = \frac{K}{4r}$ and letting $r = \text{const} \cdot m^{1/3}$ bins, the error probability of each bit is $Q(\text{const}) = \text{constant}$, i.e., does not depend on $m, r$. Finally, a one-bit error in detection causes
	\begin{align}
		&\int_{\mathcal{X}} | \beta_{\mathbf{b}}(x) - \wh{\beta}_m(x) | dx \nonumber \\
		&= \| \beta_{\mathbf{b}} - \wh{\beta}_m \|_{L_1} = \frac{2A}{r} = \text{const} \cdot m^{-1/3} \label{eq:one_bit_error}
	\end{align}
	because the first half of the error interval contributes $\frac{A}{r}$ to the distance and the other half contributes $\frac{A}{r}$.
	
	Hence, the expected energy loss for an arbitrary reconstruction method can be bounded as follows: For $\beta_{\mathbf{b}} \in F$ and at $R = C_{\maxx}$
	\begin{align*}
		&\overline{\Delta}_\ener(R;\beta_{\mathbf{b}}, \wh{\beta}_m) = \mathbb{E}\left[ |B_{\beta_{\mathbf{b}}}(R) - B_{\wh{\beta}_m}(R)|\right] \\
		&= \mathbb{E}\left[ \left| \int p_X^*(x) \beta_{\mathbf{b}}(x) dx - \int p_X^*(x) \wh{\beta}_m(x) dx \right| \right] \\
		&\stackrel{(a)}{=} c \mathbb{E}\left[ \left| \int \beta_{\mathbf{b}}(x) \bm{1}_{\{  \exists i: x \in I_i \}} dx - \int \wh{\beta}_m(x) \bm{1}_{\{  \exists i: x \in I_i \}} dx \right| \right] \\
		&= c \mathbb{E}\left[ \left| \int (\beta_{\mathbf{b}}(x) - \wh{\beta}_m(x))\bm{1}_{\{  \exists i: x \in I_i \}} dx \right| \right] \\
		&\stackrel{(b)}{\ge} \text{const} \cdot \frac{A}{r} = \text{const} \cdot m^{-1/3},
	\end{align*}
	where (a) follows since $p_X^*$ is a square pulse. Also, (b) follows since 1) the one-bit error event lower bounds the expected distance, 2) the probability of the one-bit error is a constant, and 3) $p_X^*$ only filters out the first half of an interval, which contributes $\frac{A}{r}$. It completes the proof.
\end{IEEEproof}

\subsection{Information Loss}
Recall the information loss when samples are noiseless. In the previous section, the information loss asymptotically vanishes at the interior of the energy domain only, and furthermore, the rate of convergence was the same as that for energy loss when the $C_{\beta}$ curve is Lipschitz continuous. The first conclusion still holds for noisy samples, as shown in Theorem \ref{thm:delta_I_noisy} below. However, the conclusion on the convergence rate in Corollary \ref{cor:holder_noisy} is slightly different from that for the noiseless case due to the difficulty of tail probability analysis.
\begin{thm} \label{thm:delta_I_noisy}
	For any $\beta \in \BV_K[0,1]$ and $B \in [0, B_{\maxx})$,
	\begin{align*}
		\overline{\Delta}_\info(B; \beta, \wh{\beta}_m^{\wv}) \to 0 \textrm{ as } m \to \infty.
	\end{align*}
	However, $\overline{\Delta}_{\textsf{I}}(B_{\maxx};\beta, \wh{\beta}_m^{\wv})$ does not vanish with $m$ in general.
\end{thm}
\begin{IEEEproof}
	First, at $B=0$, the capacity-energy function is the same as the unconstrained capacity. In other words, $C_{\beta}(0)=C_{\wh{\beta}_m^{\wv}}(0)=C_\maxx$. Hence, $\overline{\Delta}_\info(0; \beta, \wh{\beta}_m^{\wv}) = 0$.
	
	Next, consider $B \in (0, B_\maxx)$. Using the proof of Theorem \ref{thm:delta_E_noisy} and the Markov inequality for an arbitrary $\epsilon > 0$,
	\begin{align}
		&\mathbb{P}\left[ |B_{\beta}(R) - B_{\wh{\beta}_m^{\wv}}(R)| \ge \epsilon \right] \le \frac{\mathbb{E}\left[ |B_{\beta}(R) - B_{\wh{\beta}_m^{\wv}}(R)| \right]}{\epsilon} \nonumber \\
		&= \frac{\mathcal{O}(m^{-1/3} \sqrt{\log m})}{\epsilon} =: \delta_m \to 0 ~ \text{ as } ~ m \to \infty. \label{eq:prob_bound}
	\end{align}
	In other words, with probability at least $1-\delta_m$, the maximal deliverable energy under $\wh{\beta}_m^{\wv}$ is arbitrarily close to the energy under $\beta$. Or equivalently, there exists some $B'$ such that $C_{\beta}(B) = C_{\wh{\beta}_m^{\wv}}(B')=R$ with $|B-B'| < \epsilon$ with high probability. Therefore,
	\begin{align*}
		C_{\beta}(B) - C_{\wh{\beta}_m^{\wv}}(B) = C_{\wh{\beta}_m^{\wv}}(B') - C_{\wh{\beta}_m^{\wv}}(B),
	\end{align*}
	where $B' \in (B-\epsilon, B+\epsilon)$ with probability at least $1-\delta_m$. Since $C_{\wh{\beta}_m^{\wv}}(B)$ is concave, it is continuous on the interior of the energy domain. This in turn implies that $C_{\wh{\beta}_m^{\wv}}(B') - C_{\wh{\beta}_m^{\wv}}(B) \to 0$ as $\epsilon \to 0$ and $m \to \infty$. On the other hand, $|B_{\beta}(R) - B_{\wh{\beta}_m^{\wv}}(R)| \ge \epsilon$ with vanishing probability. This event can affect $\overline{\Delta}_\info(B;\beta, \wh{\beta}_m^{\wv})$ by at most $\delta_m C_{\maxx}$, which is asymptotically negligible as well. It proves the first claim.
	
	To prove the second claim, for instance, consider a constant harvesting function $\beta(x) = M =: B_{\maxx}$. It can obviously deliver energy $B_\maxx$ for any $p_X$, thus $C_{\beta}(B_\maxx) = C_\maxx$. However, as our samples are noisy, there is a nonvanishing probability such that the largest energy the recovered harvesting function can deliver is less than $B_\maxx$. For such a case, $C_{\wh{\beta}_m^{\wv}}(B_\maxx) = 0$, thus, $C_{\beta}(B_\maxx) - C_{\wh{\beta}_m^{\wv}}(B_\maxx) = C_\maxx$ with nonvanishing probability. It implies that $\overline{\Delta}_\info(B_\maxx; \beta, \wh{\beta}_m^{\wv})$ does not vanish.
\end{IEEEproof}

For noiseless samples, Corollary \ref{cor:holder_noiseless} when $C_{\beta}(B)$ is Lipschitz continuous implies that the decay rate of the information loss is indeed identical to that of energy loss. It is then tempting to claim the same statement for the expected energy loss in the noisy setting; however, the following corollary shows a different bound on the decay rate.
\begin{cor} \label{cor:holder_noisy}
	Suppose the energy-capacity function is $\alpha$-H\"{o}lder continuous for a harvesting function $\beta \in \BV_K[0,1]$. That is, for $B_1, B_2 \in (0, B_{\maxx})$,
	\begin{align*}
		|C_{\beta}(B_1) - C_{\beta}(B_2)| \le c |B_1 - B_2|^{\alpha}.
	\end{align*}
	Then, $\overline{\Delta}_\info(B; \beta, \wh{\beta}_m^{\wv}) = \mathcal{O} \left( m^{-\frac{\alpha}{3(\alpha+1)}} (\log m)^{\frac{\alpha}{2(\alpha+1)}} \right)$ for $B \in [0, B_{\maxx})$.
\end{cor}
\begin{IEEEproof}
	First, consider the case when $B=0$. As it reduces to the unconstrained capacity, any $p_X$ is admissible. Hence, $C_{\beta}(0) = C_{\wh{\beta}_m^{\wv}}(0)$ for any $\beta, \wh{\beta}_m^{\wv}$, implying $\overline{\Delta}_\info(B; \beta, \wh{\beta}_m^{\wv})=0$.
	
	For $B \in (0, B_{\maxx})$, let $E$ be the set of events $\{|B_{\beta}(R) - B_{\wh{\beta}_m^{\wv}}(R)| \ge \epsilon\}$. From the proof of Theorem \ref{thm:delta_I_noisy}, it is known that $\mathbb{P}[E] \le \delta_m = \frac{\mathcal{O}(m^{-1/3} \sqrt{\log m})}{\epsilon}$. Then,
	\begin{align*}
		&\mathbb{E}\left[ | C_{\beta}(B) - C_{\wh{\beta}_m^{\wv}}(B) | \right] \\
		&= \mathbb{E} \left[ | C_{\beta}(B) - C_{\wh{\beta}_m^{\wv}}(B) | \cdot \bm{1}_{\{E\}} + | C_{\beta}(B) - C_{\wh{\beta}_m^{\wv}}(B) | \cdot \bm{1}_{\{E^c\}} \right] \\
		&= C_\maxx \mathbb{P}[E] + \mathbb{E} \left[| C_{\beta}(B) - C_{\beta}(B') | \cdot \bm{1}_{\{E^c\}} \right],
	\end{align*}
	for some $B' \in (B-\epsilon, B+\epsilon)$. As $C_{\beta}$ is $\alpha$-H\"{o}lder continuous,
	\begin{align*}
		&\mathbb{E}[ | C_{\beta}(B) - C_{\wh{\beta}_m^{\wv}}(B) | ] \\
		&= C_\maxx \mathbb{P}[E] + \mathbb{E}\left[ | C_{\beta}(B) - C_{\beta}(B') | \cdot \bm{1}_{\{E^c\}} \right] \\
		&= C_\maxx \mathbb{P}[E] + c \epsilon^{\alpha} \mathbb{P}[E^c] \le C_\maxx \mathbb{P}[E] + c \epsilon^{\alpha} \\
		&\le \text{const} \cdot \frac{m^{-1/3} (\log m)^{1/2}}{\epsilon} + \text{const} \cdot \epsilon^{\alpha}.
	\end{align*}
	Optimizing the bound by matching orders, we can obtain the following:
	\begin{align*}
		&\epsilon = m^{-\frac{1}{3(\alpha+1)}} (\log m)^{\frac{1}{2(\alpha+1)}}, \\
		&\mathbb{E} \left[ | C_{\beta}(B) - C_{\wh{\beta}_m^{\wv}}(B) | \right] = \mathcal{O} \left( m^{-\frac{\alpha}{3(\alpha+1)}} (\log m)^{\frac{\alpha}{2(\alpha+1)}} \right).
	\end{align*}
\end{IEEEproof}
Note that when $\alpha=1$, i.e., $C_{\beta}(B)$ is Lipschitz continuous, the information loss reduces to
\begin{align*}
	\overline{\Delta}_\info(B; \beta, \wh{\beta}_m^{\wv}) = \mathcal{O} \left( m^{-1/6} (\log m)^{1/4} \right)= \widetilde{\mathcal{O}}(m^{-1/6}).
\end{align*}
Since it is an upper bound which might be loose, it does not directly tell whether it has the same decay as energy loss or not. The proof illustrates why it is nontrivial: Recalling that $\wh{\beta}_m^{\wv}$ is random due to noisy samples, more detailed probability bound on \eqref{eq:prob_bound} is needed, which is very challenging in general.

\section{Numerical Evaluation} \label{sec:evaluation}
This section demonstrates the performance of wavelet-based transmission for real harvesting circuit measurements and compares it with spline-based \cite{SeoV2019} and deep learning-based approaches \cite{VarastehHC2020}. As we will see, $C_{\beta_m}$ curve becomes closer to the true one as the sample size grows. It is also worth noting that the unknown harvesting function can take any shape as long as it produces the same samples, and we cannot accurately estimate it from samples without an oracle. In other words, there is always a harvesting function that largely disagrees with our estimate but still agrees with the samples. This was the key motivation for our minimax theoretical performance bound, which this section aims to numerically support as well.

For the energy harvester, we use the circuit design \cite[Figure 3]{VarastehHC2020} and measurement data with the permission of authors of \cite{VarastehHC2020}. The communication channel we assumed is the standard additive white Gaussian noise (AWGN) channel, i.e., $Y_t = X_t + Z_t$ where $Z_t \stackrel{\text{i.i.d.}}{\sim} \mathcal{N}(0,1)$.\footnote{It is known that the capacity-achieving distribution of the AWGN channel with peak-power constraint is supported by a finite set \cite{Smith1969, DytsoYPS2020, VarastehRC2020}, thus violating our problem assumption. We nonetheless consider the AWGN channel for informational purposes, as our wavelet framework still works unless the discontinuities of $\beta$ co-locate with the point masses.}

\begin{figure}[t]
\centering
\subfloat[$\beta(x)$]{\includegraphics[width=3.0in]{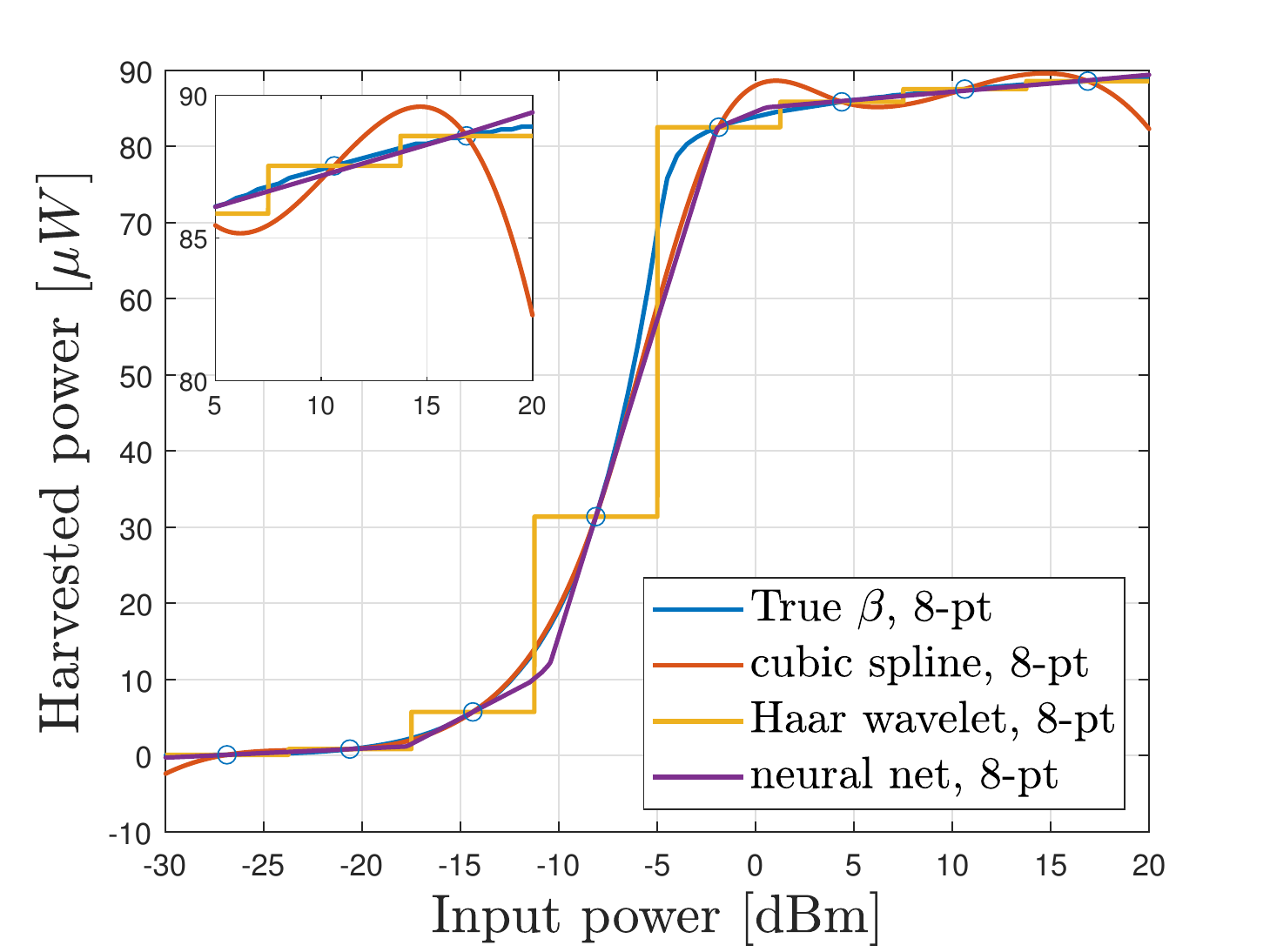}}
\caption{Function reconstruction from $8$-point samples by three methods. Since $\beta(x)$ is nearly flat near the interval's endpoints, using Haar wavelets shows the best reconstruction at endpoints.}
\label{fig:recon_ex}
\end{figure}

\begin{figure}[t]
\centering
\subfloat[$\beta(x)$]{\includegraphics[width=3.0in]{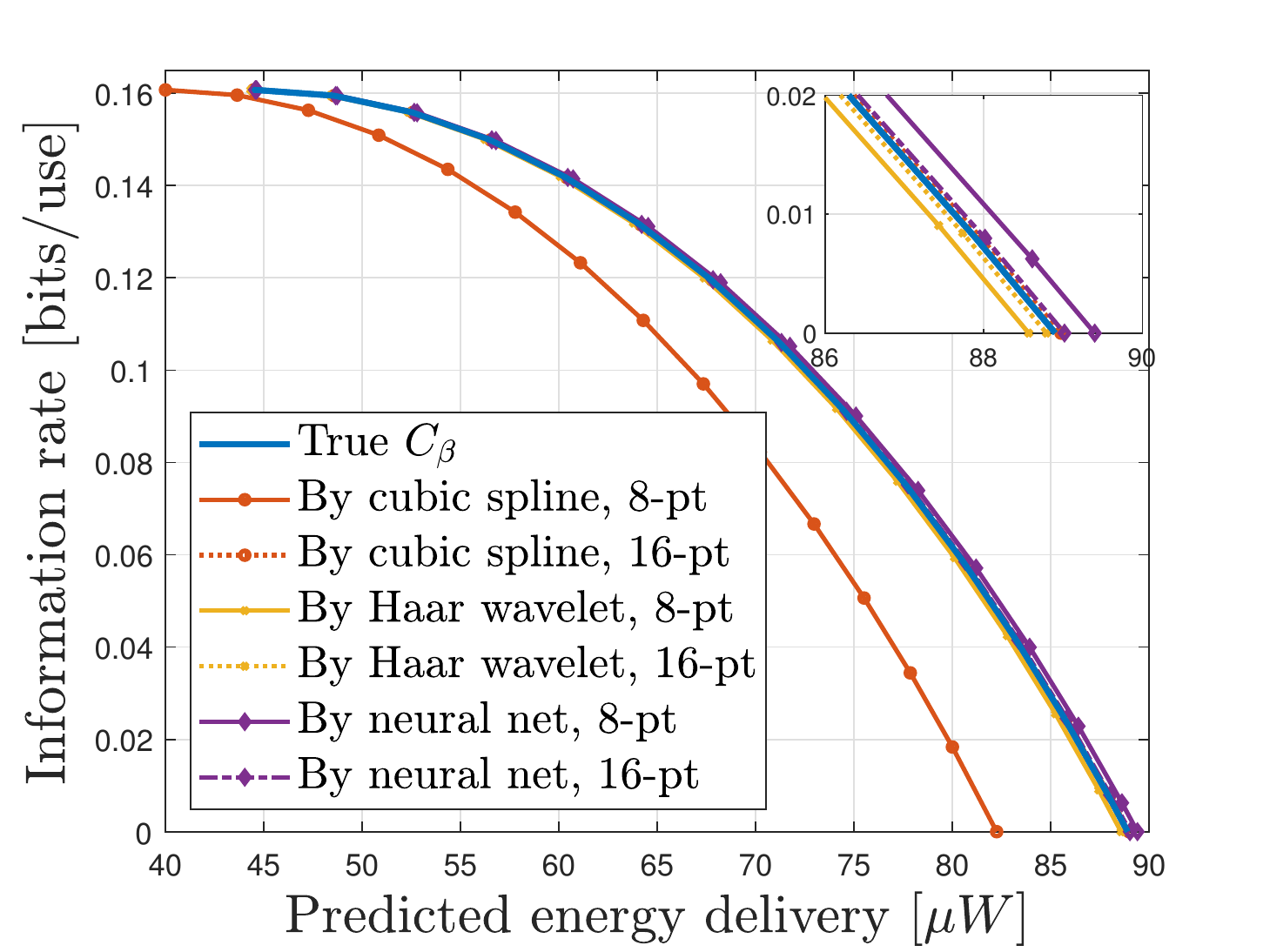}}
\caption{True $C_{\beta}$ and predicted information-energy tradeoff curves $C_{\wh{\beta}_m}$ using three methods. Note that $C_{\wh{\beta}_m}$ is the predicted energy and information tradeoff if we design SIET as if $\wh{\beta}_m$ is the truth. For both figures, more samples show better prediction. Since reconstruction by wavelets shows the smallest deviation from the truth at the endpoints (see Figure \ref{fig:recon_ex}), the wavelet method at $8$-point samples is the best prediction of the capacity curve. }
\label{fig:cap_result}
\end{figure}

The original measurements are $101$ equispaced samples. Recall that we can never know the real continuous $\beta$ from discrete samples. Hence, the $101$ samples are linearly interpolated into $2^{13}$ finer samples to model the continuous $\beta$ in $x$. This is assumed to be the true harvesting function $\beta(x)$ that is continuous in $x$. Then, we take $m=8, 16$ equispaced samples\footnote{For larger values of $m$, the results are almost indistinguishable.} in the absence of noise, from which we reconstruct $\wh{\beta}_m(x)$ using three different methods
\begin{enumerate}
	\item (Proposed method) Haar\footnote{The simplest wavelet base is used here, but it is possible to improve accuracies by selecting a different base, which can be considered as future work. } wavelets,
	\item cubic splines \cite{SeoV2019},
	\item deep learning regression that minimizes mean-squared error (MSE) \cite{VarastehHC2020},
\end{enumerate}
and then calculate their predicted Shannon capacity
\begin{align*}
	C_{\wh{\beta}_m}(B) = \max_{p_X:\mathbb{E}[\wh{\beta}_m(X)]\ge B} I(X;Y).
\end{align*}
The neural network used in the deep learning regression is a two-hidden-layer fully connected neural network, where each hidden layer has $50$ neurons with rectified linear unit (ReLU) activation. That is,
\begin{align*}
	\wh{\beta}_m(x) = W_3 \sigma( W_2 \sigma( W_1 x + b_1 ) + b_2  ) + b_3,
\end{align*}
where $\sigma$ is the ReLU activation function, and $W_i, b_i$ are respectively weight matrices and bias vectors of proper dimension. As $C_{\beta}(B)$ and $C_{\wh{\beta}_m}(B)$ are capacity-energy functions with continuous alphabet spaces, a particle-based Blahut-Arimoto algorithm \cite{Duawels2005} is implemented with fine quantization of $2^{13} = 8192$ levels. Numerical evaluations are shown in Figures \ref{fig:recon_ex} and \ref{fig:cap_result}.


Let us first discuss the $\beta$ reconstruction of the three methods in Figure \ref{fig:recon_ex}. The true $\beta$ is drawn in blue. For this particular $\beta$ with $8$ samples, in terms of average root-mean-squared error (RMSE), the best reconstruction method turns out to be the spline method with RMSE $0.0304$, while the wavelet and neural network methods show $0.0656$ and $0.0327$, respectively. However, noting that the capacity-achieving distributions of AWGN channels have discrete point masses \cite{Smith1969, DytsoYPS2020, VarastehRC2020}, which are especially supported by both endpoints in this example, reconstructing $\wh{\beta}_m$ as accurately as possible at endpoints is more critical than minimizing the average error. As $\beta(x)$ has a relatively flat region near the input interval endpoints, the use of Haar wavelets, which also have a flat behavior near endpoints, results in the most accurate reconstruction at these points, while splines exhibit the worst accuracy, as shown in the inset of Figure \ref{fig:recon_ex}. We can also propose a $\beta$ for which spline is the best and wavelet is the worst at end points---for instance, if $\beta$ fluctuates according to spline bases, then the spline will show the best accuracy even at the end points. This implies that the reconstruction accuracy highly depends on the types of splines, wavelets, and hyperparameters of neural networks such as activation functions and network architectures.

Before discussing capacity results, recall that $(B, C_{\wh{\beta}_m}(B))$ curve represents the energy and information delivery that our system inaccurately targets rather than what it can actually deliver. In other words, it only means the predicted energy and information delivery if we design SIET as if $\wh{\beta}_m$ is the truth. Therefore, our goal is to have $C_{\wh{\beta}_m}(B)$ as close as possible to $C_{\beta}(B)$ in order to minimize the deviation in performance from $C_{\beta}(B)$.

Figure \ref{fig:cap_result} illustrates how closely $C_{\wh{\beta}_m}(B)$ approximates $C_{\beta}(B)$. First, for any reconstruction method, more samples result in a closer prediction for $C_{\beta}(B)$. Larger samples $m \ge 32$ are not shown as results are almost indistinguishable. To further analyze the details, note that the capacity-achieving distribution for this example has point masses at end points, which means that the accuracy of $C_{\wh{\beta}_m}(B)$ depends only on the accuracy of $\wh{\beta}_m(x)$ at end points. As the flat bases of Haar wavelets almost perfectly reconstruct $\beta$ even with $8$ samples, $C_{\wh{\beta}_8}(B)$ by Haar wavelets is almost close to the true $C_{\beta}(B)$, while cubic splines rising/sinking near the endpoints fail to approximate $C_{\beta}(B)$ with only $8$ samples. The neural network method shows comparable performance to that of wavelets.

The numerical evaluations discussed above suggest that there is no universally good function reconstruction scheme and, thus, no universally good SIET design principle. It can be thought of as an adversarial game---once we attempt to reconstruct $\beta(x)$ from samples as accurately as possible, an adversary can always choose a function that largely disagrees with our estimate but agrees with sample points. Since the true harvesting function is unavailable, it is impossible to know whether the true function is favorable or not with our scheme. Therefore, it justifies our theoretical performance guarantee on the worst-case definitions over all possible functions in the function class.

\section{Conclusion} \label{sec:conclusion}
In this work, we have studied SIET assuming that the harvesting function is of bounded variation and partially known via experimental samples in the absence or in the presence of noise. We propose to use wavelet reconstruction and design SIET based on it. The performance of such design is evaluated under the (expected for noisy samples) loss of energy and information rate. For both cases, we observe that the (expected) energy and information losses are asymptotically negligible if a target energy level is in the interior of the deliverable energy domain. For noisy samples, the obtained expected energy loss turns out to be asymptotically tight up to a logarithmic factor. We also numerically evaluate the proposed scheme, which confirms that more samples approximate the optimal SIET system more accurately and justifies our minimax approach with theoretical guarantees.

The proposed wavelet-based SIET has several future works. Note that our SIET design is sequential: It first reconstructs the harvesting function using wavelets and then designs Shannon's random codebook that achieves the capacity-energy function. The same principle can be adopted for another objective of SIET. For instance, the signal constellation for digital communication \cite{VarastehHC2020} or waveform design \cite{ClerckxB2016} with/without channel state information \cite{JoudehC2016} can also be optimized on top of the wavelet-reconstructed harvesting function. Also, extending the current results to a multi-dimensional input, e.g., multiple-input and multiple-output (MIMO) transmission, and an energy harvester with memory would be interesting.

\section*{Acknowledgement}
We thank Hansung Choi for his help with deep learning simulations in Section \ref{sec:evaluation}.

\appendices
\section{Proof of Proposition \ref{prop:noiseless}} \label{app:pf_prop1}
Note that $\cup_j V_{j}$ is dense in $L_2[0,1]$, that is, any $f \in L_2[0,1]$ can be accurately recovered by infinitely fine wavelets. As wavelets are orthogonal to each other, the error due to the finite scale $J$ is
\begin{align}
	\|f-\wh{f}_J^{\wv}\|_{L_2}^2 &= \left\| \sum_{j \ge J+1} \sum_{k=0}^{2^j-1} \langle f, \psi_{j,k} \rangle \psi_{j,k} \right\|_{L_2}^{2} \nonumber \\
	&= \sum_{j \ge J+1} \sum_{k=0}^{2^j-1} | \langle f, \psi_{j,k} \rangle |^2, \label{eq:null_component}
\end{align}
as $\psi_{j,k}$ are orthogonal each other. Furthermore, letting $\Psi(x)$ be the antiderivative of $\psi(x)$, 
\begin{align*}
	| \langle f, \psi_{j,k} \rangle | &= \left| \int_{0}^1 f(x) 2^{j/2} \psi(2^j x - k) dx \right| \\
	&\stackrel{(a)}{=} \left| \int_{0}^1 f^{(1)}(x) 2^{-j/2} \Psi(2^j x - k) dx \right| \\
	&= 2^{-j/2} \Bigg| \int_{\text{supp}(\Psi(2^j x - k))} f^{(1)}(x) \Psi(2^j x - k) dx  \Bigg| \\
	&\le 2^{-j/2} \max_x |\Psi(x)| \int_{\text{supp}(\Psi(2^j x - k))} |f^{(1)}(x)| dx \\
	&\le C_{\Psi}  2^{-j/2} \int_{I_{j,k}} |f^{(1)}(x)| dx,
\end{align*}
where (a) follows from integration by parts, and $C_{\Psi} := \max_x |\Psi(x)|$ and $I_{j,k} := \text{supp}(\Psi(2^j x - k))$. Using the fact that the $\ell_2$-norm is smaller than the $\ell_1$-norm,
\begin{align*}
	&\left( \sum_{k=0}^{2^j-1} | \langle f, \psi_{j,k} \rangle |^2 \right)^{1/2} \le \sum_{k=0}^{2^j-1} | \langle f, \psi_{j,k} \rangle | \\
	&\le \sum_{k=0}^{2^j-1} C_{\Psi}  2^{-j/2} \int_{I_{j,k}} |f^{(1)}(x)| dx \le C_{\Psi} C_j 2^{-j/2} \TV(f),
\end{align*}
where $C_j$ is the multiplicity taking the overlaps in $I_{j,k}$'s into account, and $C_j \le C'$ for some $C'$ since it is determined by $\psi(x)$. Then, we can obtain
\begin{align}
	&\|f - \wh{f}_J^{\wv}\|_{L_2}^2 = \sum_{j \ge J+1} \sum_{k=0}^{2^j-1} | \langle f, \psi_{j,k} \rangle |^2 \nonumber \\
	&\le \sum_{j \ge J+1} (C_{\Psi} C')^2 2^{-j} \TV^2(f) \label{eq:coeff_magnitude} \\
	&= 2^{-J}(C_{\Psi} C')^2 \TV^2(f) \nonumber \\
	&= c \cdot \TV^2(f) \cdot m^{-1} ~~~\text{with } c:= (C_{\Psi}C')^2. \nonumber
\end{align}
The statement is proved.

\section{Proof of Proposition \ref{prop:noisy_wavelet_error}} \label{app:pf_prop2}
Before proceeding, we state two building blocks. The first lemma approximates a function by a wavelet series having $t$ terms, and the second lemma estimates coefficients from noisy samples.
\begin{lem} \label{lem:best_m_term_approx}
	Let $f \in \BV_K[0,1]$ be a function to be recovered. Suppose that the wavelet has at least one vanishing moment and a compact support. Then, we can take a $t$-term wavelet series $f_t$ that satisfies
	\begin{align*}
		\|f-f_t \|_{L_2}^2  = \mathcal{O}(t^{-2}).
	\end{align*}
\end{lem}
\begin{IEEEproof}
	From \eqref{eq:coeff_magnitude}, we know that at each scale $j$,
	\begin{align*}
		\sum_{k=0}^{2^j-1} |\theta_{j,k}| \le c 2^{-j/2},
	\end{align*}
	where $\theta_{j,k} := \langle f, \psi_{j,k} \rangle$. If we reorder the coefficients such that $|\theta_{j,(0)}| \ge |\theta_{j,(1)}| \ge \cdots \ge |\theta_{j,(2^{j}-1)}|$, we can obtain
	\begin{align*}
		c 2^{-j/2} \ge \sum_{k=0}^{2^j-1} |\theta_{j,(k)}| \ge \sum_{k=0}^{r} |\theta_{j,(r)}| \ge (r+1) |\theta_{j,(r)}|
	\end{align*}
	for any $r \in \{0, \ldots, 2^j-1\}$. This implies that $|\theta_{j,(r)}| \le c 2^{-j/2} (r+1)^{-1}$.
	
	If a threshold $T$ is taken, then the number of coefficients having a magnitude greater than $T$ at scale $j$, say $n_j(T)$, is
	\begin{align*}
		n_j(T) \le \min\{ c 2^{-j/2} T^{-1}, 2^{j} \}.
	\end{align*}
	Therefore, the total number of coefficients having a magnitude greater than $T$ at all scales is
	\begin{align*}
		n(T) &\le \sum_{j=0}^{\infty} \min\{ c 2^{-j/2} T^{-1}, 2^{j} \} \\
		&\stackrel{(a)}{=} \sum_{j \le j^*} 2^j + \sum_{j > j^*} c 2^{-j/2} T^{-1} \\
		&\stackrel{(b)}{=} \mathcal{O}(2^{j^*}) + \mathcal{O}(2^{-j^*/2} T^{-1}) \\
		&= \mathcal{O}(T^{-2/3}) + \mathcal{O}(T^{-2/3}) = \mathcal{O}(T^{-2/3}).
	\end{align*}
	where (a) follows since $2^j = c 2^{-j/2} T^{-1}$ at $j^* = \frac{2}{3} \log_2 cT^{-1}$ and (b) follows from the sum of a geometric series.
	
	Let $\theta_{(r)}$ be the $r$-th largest coefficient over all scales. Then, we have
	\begin{align*}
		|\theta_{(r)}| \le \widetilde{c} r^{-3/2}.
	\end{align*}
	It implies that by choosing $t$ largest coefficients over all scales, we can construct $f_t$ such that 
	\begin{align*}
		\| f - f_t \|_{L_2}^2 &= \sum_{k=t+1}^{\infty} |\theta_{(k)}|^2 \\
		&\stackrel{(c)}{=} \sum_{k=t+1}^{\infty} \widetilde{c} k^{-3} = \mathcal{O}(t^{-2}).
	\end{align*}
	where (c) follows from the Parseval–Plancherel theorem. The claim is proved.
\end{IEEEproof}

\begin{lem}[Theorem 1 in \cite{DonohoJ1994}] \label{lem:donoho_johnstone}
	Let $\bm{\theta} = (\theta_1, \ldots, \theta_p) \in \mathbb{R}^p$ be a vector to be estimated. Suppose that we have noisy observations
	\begin{align*}
		Y_i = \theta_i + Z_i, ~~ Z_i \stackrel{\text{i.i.d.}}{\sim} N(0, \sigma^2).
	\end{align*}
	Then, the soft-thresholding estimator $\wh{\theta}_i = \sign(y_i) \max(|y_i| - \lambda, 0)$ with $\lambda = \sqrt{2\sigma^2 \log p}$ achieves the expected error
	\begin{align*}
		\mathbb{E}\left[ \|\bm{\theta} - \wh{\bm{\theta}} \|_2^2 \right] \le (2\log p + 1) \left( \sigma^2 + \sum_{i=1}^p \min(\theta_i^2, \sigma^2) \right).
	\end{align*}
\end{lem}

\begin{IEEEproof}[Proof of Proposition \ref{prop:noisy_wavelet_error}]
	We will first project $f$ into $V_J$ space (approximation) and then estimate $\Proj_{V_J}(f)$ from noisy samples (estimation). For wavelet coefficient estimation, soft thresholding will be used.
	
	Note that $f-\Proj_{V_J}(f)$ and $\Proj_{V_J}(f) - \wh{f}^{\wv}$ are orthogonal since $\Proj_{V_J}(f) - \wh{f}^{\wv}$ belongs to $V_J$. That is,
	\begin{align*}
		&\mathbb{E}\left[ \|f-\wh{f}^{\wv}\|_{L_2}^2 \right] \\
		&= \mathbb{E} \left[ \|f- \Proj_{V_J} (f)\|_{L_2}^2 \right] + \mathbb{E}\left[ \|\Proj_{V_J} (f) - \wh{f}^{\wv}\|_{L_2}^2 \right] \\
		&= \|f- \Proj_{V_J} (f)\|_{L_2}^2 + \mathbb{E}\left[ \|\Proj_{V_J} (f) - \wh{f}^{\wv}\|_{L_2}^2 \right],
	\end{align*}
	where the first term has no expectation operator since it has nothing to do with random noise. Further, the first term is $\mathcal{O}(m^{-1})$ by Proposition \ref{prop:noiseless}. We will see that this is asymptotically negligible compared to the second term.
	
	To obtain a bound on the second term, first, consider a na\"{i}vely reconstructed function $\wh{g}$: Let $f_Y(x)$ be a piecewise constant function obtained from samples. Then, $\wh{g}(x) = \sum_{j=0}^{J-1} \sum_{k=0}^{2^j-1} \wh{c}_{j,k} \psi_{j,k}(x) = \sum_{j=0}^{J-1} \sum_{k=0}^{2^j-1} \wh{c}_{j,k} 2^{j/2} \psi(2^jx -k)$ with coefficients 
	\begin{align*}
		\wh{c}_{j,k} &= \langle f_Y, \psi_{j,k} \rangle = \int f_Y(x) \psi_{j,k}(x) dx \\
		&= \int f_Y(x) 2^{j/2}\psi(2^jx-k) dx \\
		&= 2^{-j/2} \int f_Y(2^{-j}z) \psi(z-k) dz = c_{j,k} + \widetilde{Z}_{j,k},
	\end{align*}
	where $c_{j,k}$ is the true coefficient, and $\widetilde{Z}_{j,k}$ is the equivalent Gaussian noise with variance $\widetilde{\sigma}^2 = \text{const}\cdot 2^{-J} \sigma^2 \propto m^{-1}$.
	
	Let $\bm{\theta}$ be the vectorized version of $\{c_{j,k}\}$. Then, instead of $\theta_i$, we only have noisy coefficients corrupted by $\widetilde{Z}_i$. Recall that $\| \Proj_{V_J} (f) - \wh{f}^{\wv} \|_{L_2}^2$ is the same as coefficient estimation error as wavelets are orthogonal. Therefore, estimating the coefficients using soft thresholding in Lemma \ref{lem:donoho_johnstone} gives
	\begin{align*}
		&\mathbb{E}\left[ \|f - \wh{f}^{\wv}\|_{L_2}^2 \right] = \mathcal{O}(m^{-1}) + \mathbb{E}\left[ \|\Proj_{V_J} (f) - \wh{f}^{\wv}\|_{L_2}^2 \right] \\
		&\le \mathcal{O}(m^{-1}) + (2 \log m + 1) \left( \widetilde{\sigma}^2 + \sum_{i=1}^m \min(\theta_i^2, \widetilde{\sigma}^2) \right).
	\end{align*}
	
	We can further upper bound it via elaboration on the last term using Lemma \ref{lem:best_m_term_approx}. Let $\theta_{(0)} > \theta_{(1)} > \cdots > \theta_{(m-1)}$ be the ordered coefficients. Then, using the property that $\min(x,y) \le x$ and $\min(x,y) \le y$,
	\begin{align*}
		\sum_{i=0}^{m-1} \min(|\theta_{(i)}|^2, \widetilde{\sigma}^2) &\le \sum_{i=0}^{t} \widetilde{\sigma}^2 + \sum_{i=t+1}^{m-1} \theta_{(i)}^2 = t \widetilde{\sigma}^2 + \mathcal{O}(t^{-2}),
	\end{align*}
	where the last term follows from Lemma \ref{lem:best_m_term_approx}. Recalling that $\widetilde{\sigma}^2 = \mathcal{O}(m^{-1})$ and matching the speed of decay, we get $t^* = m^{1/3}$. It, in turn, gives us
	\begin{align*}
		\mathbb{E}\left[ \|f-\wh{f}^{\wv}\|_{L_2}^2 \right] &\le \mathcal{O}(m^{-1}) + (2\log m + 1) (\text{const} \cdot m^{-2/3}) \\
		&= \mathcal{O}(m^{-2/3} \log m).
	\end{align*}
\end{IEEEproof}

\begin{IEEEbiographynophoto} {Daewon~Seo} received the B.S.~(summa cum laude) and M.S.~degrees in electrical engineering from the Korea Advanced Institute of Science and Technology (KAIST), Daejeon, South Korea, in 2008 and 2010, respectively. After the M.S.~degree, he was with KAIST Institute and LG Electronics, South Korea, from 2010 to 2011 and from 2011 to 2014, respectively. He received the Ph.D.~degree in electrical and computer engineering from the University of Illinois Urbana-Champaign, Champaign, IL, USA in 2019. He was with the Ming Hsieh department of electrical and computer engineering, the University of Southern California, Los Angeles, CA, USA, in 2019 and the department of electrical and computer engineering, the University of Wisconsin-Madison, Madison, WI, USA from 2019 to 2020. He is currently an assistant professor in the department of electrical engineering and computer science at Daegu Gyeongbuk Institute of Science and Technology (DGIST), Daegu, South Korea. His research interests include information theory, statistical inference, and machine learning systems.
\end{IEEEbiographynophoto}

\begin{IEEEbiographynophoto}
{Yongjune Kim} (Member, IEEE) received the B.S. (Hons.) and M.S. degrees in electrical and computer engineering from Seoul National University, Seoul, South Korea, in 2002 and 2004, respectively, and the Ph.D. degree in electrical and computer engineering from Carnegie Mellon University, Pittsburgh, PA, USA, in 2016. From 2016 to 2018, he was a Postdoctoral Research Associate in the Coordinated Science Laboratory at University of Illinois at Urbana-Champaign, Urbana, IL, USA. From 2018 to 2020, he was with Western Digital Research, Milpitas, CA, USA. From 2020 to 2022, he was an Assistant Professor in the Department of Electrical Engineering and Computer Science, Daegu Gyeongbuk Institute of Science and Technology (DGIST), Daegu, South Korea. Since 2022, he has been with the Department of Electrical and Engineering, Pohang University of Science and Technology (POSTECH), Pohang, South Korea, where he is currently an Assistant Professor. He is also an Adjunct Professor at Yonsei University, Seoul, South Korea. His research interests include semantic communications, coding and information theory, and machine learning. He received the IEEE Data Storage Best Student Paper Award, the Best Paper Award of the 2016 IEEE International Conference on Communications (ICC), the Best Paper Award (honorable mention) of the 2018 IEEE International Symposium on Circuits and Systems (ISCAS), and the Best Paper Award of the 31st Samsung Semiconductor Technology Symposium.
\end{IEEEbiographynophoto}

\vfill

\end{document}